\documentclass[a4paper,11pt]{article}
\pdfoutput=1 % if your are submitting a pdflatex (i.e. if you have
\usepackage{jcappub} % for details on the use of the package, please
\usepackage{txfonts}
\usepackage{lineno}

\usepackage[T1]{fontenc} % if needed
\usepackage{amsmath}	% Advanced maths commands
\usepackage{amssymb}	% Extra maths symbols

\usepackage{graphicx}	% Including figure files
\usepackage{subfigure}

\title{\boldmath Negative Turbulent Magnetic Diffusivity $\beta$ effect in a Magnetically Forced System}

\author[a,1]{Kiwan Park,\note{Corresponding author.}}
\author[a]{Myung-Ki Cheoun}
\affiliation[a]{Department of Physics and Origin of Matter and Evolution of Galaxy (OMEG) Institute, Soongsil University, Seoul 156-743, Korea}
\emailAdd{pkiwan@ssu.ac.kr}
\abstract
{\\\\
We have studied the large scale dynamo process forced with helical magnetic energy (magnetic helicity). The magnetically driven dynamo is not so well studied as kinetically forced dynamo. It has been thought to represent the amplification of magnetic field in the stellar corona, accretion disk, or plasma lab. However, the interaction between the helical magnetic field and plasma is a more fundamental phenomenon that can be extended to the early Universe. The scale-invariant helical magnetic field not only explains the currently observed large scale astrophysical magnetic fields but also has information on the horizon scale in the early Universe.\\

\noindent The interaction between magnetic field and plasma is inherently non-linear, making its mechanism difficult to understand. But, if the plasma system is driven with helical field, the process can be linearized with $\alpha$\&$\beta$ and large scale magnetic field ${\overline{\bf B}}$. Conventionally, $\alpha$ effect is thought to transfer magnetic field to the large scale regime, and $\beta$ effect diffuses magnetic field. However, these conclusions are based on the incompletely derived $\alpha$\&$\beta$. In this paper, to get the exact profiles of evolving $\alpha$\&$\beta$, we solved a coupled semi-analytic equation set and applied the result to simulation data for the large scale magnetic helicity $\overline{H}_M$ and magnetic energy $\overline{E}_M$.\\

\noindent Our result shows that the averaged $\alpha$ effect decreases before making a significant contribution to the amplification of ${\overline{\bf B}}$ field. Rather, $\beta$ effect, which keeps negative, de facto plays a key role in the amplification of $\overline{B}$ field with Laplacian ($\nabla^2\rightarrow -k^2$). And, this negative diffusivity accounts for the attenuation of plasma kinetic energy $\overline{E}_V$. Helical plasma velocity field $U$ plays a more complex role in dynamo. In addition to the conventional diffusion effect, poloidal field ${\bf U}_{pol}$ and toroidal field ${\bf U}_{tor}$ interact with ${\bf B}\cdot\nabla {\bf U}$ and $-{\bf U}\cdot\nabla {\bf B}$ to produce $\alpha$ effect and negative $\beta$ effect. We discussed this process using the theoretical method and the intuitive field structure model.}

\begin{document}
\maketitle
\flushbottom

\section{Introduction and method}
Most celestial plasma systems are constrained by magnetic field $B$. $B$ field takes energy from the turbulent plasma (dynamo), and the amplified field back reacts to the system (magnetic back reaction). Through this mutual interaction, $B$ field controls the rate of formation of a star and accretion disk \citep{Balbus, Machida}. Also, the balanced pressure between the magnetic field and plasma can decide the stability of the system (see sausage, kink, or Kruskal-Schwarzschild instability, see \cite{Boyd}). {However, their detailed internal mechanisms related to the interaction between plasma and magnetic field are not yet clearly understood.}\\

\noindent The amplification of $B$ field in plasma usually requires seed magnetic field. However, the origin of seed field (primordial magnetic field, PMF) is still under debate. At present, its cosmological origins are divided into the era of inflationary genesis and post-inflationary magneto-genesis. \\

\noindent The first inflationary scenario generates the very large scale PMF, but it needs the breaking of conformal symmetry by the interaction of the electromagnetic field and the gravitational field. The breaking of the conformal symmetry is to consider the Electro-Magnetic (EM) coupling to scalar field \citep{Martin, Subramanian}, coupling to the modified general relativity f(R) theory, coupling to pseudo scalar field and so on. The PMF strength could be generated by quantum fluctuations and has been estimated as $10^{-5}nG - 1nG$\citep{Yamazaki}.\\

\noindent The second scenario is based on the cosmological Quantum Chromo Dynamics (QCD) phase transition ($\sim$250MeV) \citep{Cheng, Tevzadze} and the electroweak phase transition ($\sim$100MeV). The PMF could be generated by collision and percolation of some bubbles from the first order transition and estimated as $10^{-7}$nG by the quark-hadron and $10^{-14}$nG - $10^{-8}$nG order by the electroweak transition.\\

\noindent The third scenario can occur during or after the epoch of photon last scattering. The PMF can be  produced by non-vanishing vorticity, which arises from the non-zero electron and proton fluid angular velocities by the different masses of proton and electron in the gravitational field (Harrison’s mechanism, \cite{Harrison}). The PMF is thought to be about $10^{-9} nG$\\

\noindent The second and third scenarios are thought to occur on a correlation scale smaller than the Hubble radius, by which we expect a suitable field generated by another dynamical effect, for instance, Biermann battery mechanism \citep{Biermann}. When the hot ionized particles (plasma) collide mutually, the fluctuating electron density $\nabla n_e$ and pressure $\nabla p_e$ (or temperature $\nabla T_e$) can be misaligned. This instability $-\nabla p_e/n_ee$ can drive currents to generate magnetic fields. Also, the neutrino interaction with charged leptons at the early epoch is thought to have generated primordial magnetic helicity, the measure of twist and linkage of magnetic fields ($H_M=\langle {\bf A}\cdot {\bf B}\rangle$, ${\bf B}=\nabla \times {\bf A}$) \citep{Semikoz_Sokoloff}. But, since the neutrino interaction exists not only in the early universe epoch but also are abundant in the present Universe including the Sun, lepton-neutrino interaction in primordial plasma can be one of the promising candidates of (origin) magnetic field generation.\\

{\bf \noindent The PMF may have scalar, vector and tensor fluctuations, which may affect CMB observables such as the CMA anisotropies and the matter power spectrum. Also the helical PMF may produce parity-odd cross correlations which results in the non-Gaussian CMB. The PMF effect on the BBN has been shown to be not large enough to explain the Li7 problem \citep{Yamazaki et al}. But a recent paper \citep{Luo et al} assumed inhomogeneous PMF, which causes inhomogeneous temperature and non-Maxwell Boltzmann (MB) velocity distributions of baryons, and affect the relevant nuclear reaction on the BBN epoch.\\}

{\bf \noindent These quantum fluctuating effects on magnetic field generation were followed by that of the collective plasma particles, i.e., fluid. The statistical feature of charged particles constraining one another became more important.} Their aggregative motion formed a flow, which interacted with the seed magnetic fields, led to the amplification (dynamo) of $B$ field or its decay according to the various conditions. The evolution of magnetic field is now explained with Faraday equation ${\partial \bf B}/\partial t=-\nabla \times  {\bf E}$ combined with Ohm's law $\eta{\bf J}=({\bf E}+{\bf U}\times {\bf B})$ in the level of magnetohydrodynamics (MHD).\footnote{${\bf J},\, \eta,\, {\bf E},\, {\bf U}$ are current density, magnetic diffusivity, electric field, and plasma velocity.} This combined equation, i.e., magnetic induction equation implies that any electromagnetic instability such as Biermann's battery effect or lepton-neutrino interaction can be merged into electromotive force (EMF, $\sim{\bf U}\times {\bf B}$ omitting $\int d\tau$) in the equation: ${\partial \bf B}/\partial t=\nabla\times ({\bf U}\times {\bf B}-\eta{\bf J}+{\bf f}_{mag})$. Then, the provided magnetic energy grows and propagates in the plasma system. This type of process is called magnetic forcing dynamo (MFD).\\

\noindent In comparison with kinetic forcing dynamo (KFD), MFD has its own characteristic features. For example, seed magnetic field is not necessary, and magnetic helicity is not conserved. The most unusual thing is that plasma kinetic energy ($E_V=U^2/2$) is not converted into magnetic energy ($E_M=B^2/2$). Rather, partial magnetic energy is converted into kinetic energy through Lorentz force. The converted velocity field forms EMF, which transfers magnetic energy. This energy convert process minimizes the thermal dissipation in plasma so that the amplification of magnetic field in MFD is more efficient than that of KFD. In our case, we used only 20\% of forcing strength compared to KFD, but the saturated magnetic energy is larger. MFD itself is one of the prominent processes that connect the cosmological magnetic field and the seed magnetic field for KFD.\\

%\noindent The induction, amplification, and propagation of $B$ field in plasma are restricted by the massive particle motion through EMF and dissipation effect. So, dynamo in plasma is not guaranteed. Moreover, without some specific conditions, $B$ field coupled with the massive charged particles usually cascades toward the smaller scale regime and finally disappears at the dissipation scale (small scale dynamo, SSD). In contrast, with some specific condition such as helicity, shear, or instability, the field can be transferred to the larger scale (large scale dynamo, LSD, see \citep{Krause, Moffatt1978, Brandenburg_Subramanian, Balbus, Park2012a}). These features distinguish the magnetic field in plasma from the one as an electromagnetic wave in free space.\\

\noindent Basically, dynamo is a nonlinear phenomenon. However, when the field is helical {\bf ($\nabla \times {\bf F}=\lambda {\bf F}$)}, the dynamo process, i.e., EMF can be linearized with $\alpha$ \& $\beta$ and large scale magnetic field $\overline{\bf B}$. At present, the exact analytic forms of $\alpha$ \& $\beta$ are unknown. Only, their sketchy representations can be derived with closure theory and function reiterative method, e.g., mean field theory (MFT, \cite{Moffatt1978}), eddy damped quasi normal markovianized approximation (EDQNM, \cite{Pouquet}), direct interactive approximation (DIA, \cite{Yoshizawa}). Physically, $\alpha$ effect is thought to arise with Coriolis force and buoyancy  (kinetic helicity `$-\langle {\bf u}\cdot \nabla \times {\bf u}\rangle$) and gradually becomes quenched by current helicity `$\langle {\bf b}\cdot \nabla \times {\bf b}\rangle\,(=k^2\langle {\bf a}\cdot \nabla \times {\bf b}\rangle$, k: wavenumber in Fourier space)' generated by the growing magnetic back reaction\footnote{The over bar in the variable $\overline{X}$ means the large scale quantity, and the small letter $x$ means the quantity in the small (turbulent) scale regime. And, the angle bracket indicates its spatial average over the large scale regime: $(1/2L)^3 \int^L_{-L} {X} d{\bf r}$. We assume $\langle \overline{X} \rangle\sim \overline{X}$}. Adding these two effects qualitatively explains how the large scale magnetic field grows and finally becomes saturated. However, for some systems like Solar(stellar) corona or a jet structure above the accretion disk, it is hard to expect that such helical kinetic motion exists and triggers the dynamo process. Rather, the transferred helical magnetic field  {\bf $(\nabla \times {\bf B}=\lambda {\bf B}$)} from the structures is more likely to play a key role in dynamo.\\

%For example, in a rotating spherical plasma system such as the Sun, Coriolis force and buoyancy form a helically structured motion, i.e.,  kinetic helicity $\langle {\bf U}\cdot \nabla \times {\bf U}\rangle$. This yields the helical magnetic structure (current helicity $\langle {\bf J}\cdot {\bf B}\rangle$, ${\bf J}=\nabla \times {\bf B}$) with opposite chirality. Kinetic helicity and current helicity in small scale regime compose $\alpha$ effect; and the turbulent plasma motion produces $\beta$ effect.\\

%$\sim  \int^t (\langle {\bf b}\cdot \nabla \times {\bf b}\rangle-\langle {\bf u}\cdot \nabla \times {\bf u}\rangle)d\tau$  (${\bf x}={\bf X}-\overline{\bf X}$, $\overline{\bf X}$: large scale value).

% ($\sim \int^t \langle u^2 \rangle d\tau$)

%%%%%%%% $\partial {\bf u}/\partial t$ yields $\langle {\bf b}\cdot \nabla \times {\bf b}\rangle$, and $\partial {\bf b}/\partial t$ yields $-\langle {\bf u}\cdot \nabla \times {\bf u}\rangle$ and magnetic diffusivity.

\noindent Theoretically, $\alpha$ effect is derived from the differentiation of EMF over time: $\partial\langle{\bf u}\times {\bf b}\rangle/\partial t$. $\partial {\bf u}/\partial t$ and $\partial {\bf b}/\partial t$ are replaced by MHD equations leading to $\alpha\sim \int dt (-\langle {\bf u}\cdot \nabla \times {\bf u}\rangle+\langle {\bf b}\cdot \nabla \times {\bf b}\rangle)$. During this analytic calculation, there is no constraint on the derivation of $-\langle {\bf u}\cdot \nabla \times {\bf u}\rangle$ and $\langle {\bf b}\cdot \nabla \times {\bf b}\rangle$. The calculation is done only with tensor identity and the assumption of isotropy. This implies that we can drive the system either kinetic energy or magnetic energy. Current density $\bf J$ is the source of magnetic field as Biot-Savart law and Maxwell theory indicate. Moreover, to produce magnetic helicity in a lab., $\bf J$ is transmitted along magnetic field. These theoretical and experimental examples show that helical magnetic forcing dynamo (HMFD) is not forbidden but a feasible process in nature.\\

\noindent However, there are a couple of things to be made clear in HMFD. The growth rate ($\alpha$) should be larger than the dissipation rate to arise magnetic field. If current helicity is a unique component in $\alpha$ effect, magnetic field grows without stop. To prevent this catastrophic amplification, there should be sort of a constraining effect such as kinetic helicity. However, since helical magnetic field $\nabla \times {\bf B}\sim {\bf B}$ nullifies Lorentz force ${\bf J}\times {\bf B}$, the generation of helical velocity field by ${\bf B}$ looks contradictory. Moreover, even if the generation of $\bf U$ is  explained, there remain tricky issues in the conservation and chirality of  helicity. For example, if the system is forced by right handed($+$) helical kinetic energy, left handed($-$) magnetic helicity is generated and inversely cascaded. And, this ($-$) magnetic helicity in the large scale produces ($+$) magnetic helicity in the small-scale regime to conserve magnetic helicity in the system. However, in HMFD, if ($+$) helical magnetic energy drives the system, ($+$) magnetic helicity is generated in the whole scale.\\

\noindent {\bf Besides, the scale invariant helical magnetic field, i.e., magnetic helicity provides us with information on the horizon during inflation. The correlation length of PMF is constrained by hubble horizon, beyond which the correlation vanishes with sudden cut-off. Therefore, within the subhorizon with the condition of ${\bf B}\cdot {\bf \hat{n}}=0$, magnetic helicity becomes gauge free. However, the correlation length scale will be smaller than the typical galaxy scale. Even if the scale had been expanded through MHD process, the strength of magnetic field would have become too weak. An efficient dynamo process by helicity leading to both inverse cascade of magnetic energy and scale expansion is required.\\}

%\noindent {\bf By the way, magnetic helicity is not just for the convenience of calculation. The helical magnetic field explains the large scale dynamo in the Universe. Also, the scale invariant magnetic helicity produced during inflation may have information on the horizon at that time. The correlation length of PMF is constrained by hubble horizon. Beyond the horizon, the correlation vanishes or shows sudden cut-off. Within the subhorizon with the condition of ${\bf B}\cdot {\bf \hat{n}}=0$, magnetic helicity becomes gauge free. However, the correlation length scale will be smaller than the typical galaxy scale. Even if the scale expands through MHD process, the strength of magnetic field will become too weak. Therefore, an efficient dynamo process by helicity causing both inverse cascade of magnetic energy and the scale expansion is required.\\}

\noindent In the present paper, to explain the inconsistent features depending on the forcing method, we apply the analytic method and field structure model used in helical kinetic forcing dynamo \citep{Park2020} to helical magnetic forcing dynamo. Followed by this introductory section, numerical method and related MHD equations for simulation are introduced in chapter 2. In chapter 3, we show numerical results for the evolving ${B}$ field and its inverse cascade to the large scale regime. And then, we show the evolving profile of $\alpha$ \& $\beta$ along with the growth of $B$ field and investigate their physical features and mutual relations. In chapter 4, we discuss the parameterisations of EMF: $\langle {\bf u}\times {\bf b}\rangle \sim \int d\tau (\alpha \overline{\bf B} -\beta \nabla \times \overline{\bf B}$) and compare them using numerical data and analytic approach. Using field structure model, we explain the intuitive meaning of $\alpha$ effect and how $\beta$ becomes negative. Then, to supplement them, we derive $\beta$ coefficient again when the field is helical. $\beta$ effect also explains how the plasma velocity field is suppressed when the system is forced by helical magnetic field. This work focuses on the physical mechanism of helical forcing dynamo which occurs in the fundamental level of astro-plasma system. {In chapter 5, we summarize our work.}\\

%In contrast, the positive $\beta$ from nonhelical turbulent plasma energy diffuses magnetic energy into the large scale plasma motion.\\

\begin{figure*}
\centering{
   \subfigure[]{
     \includegraphics[width=7.5 cm]{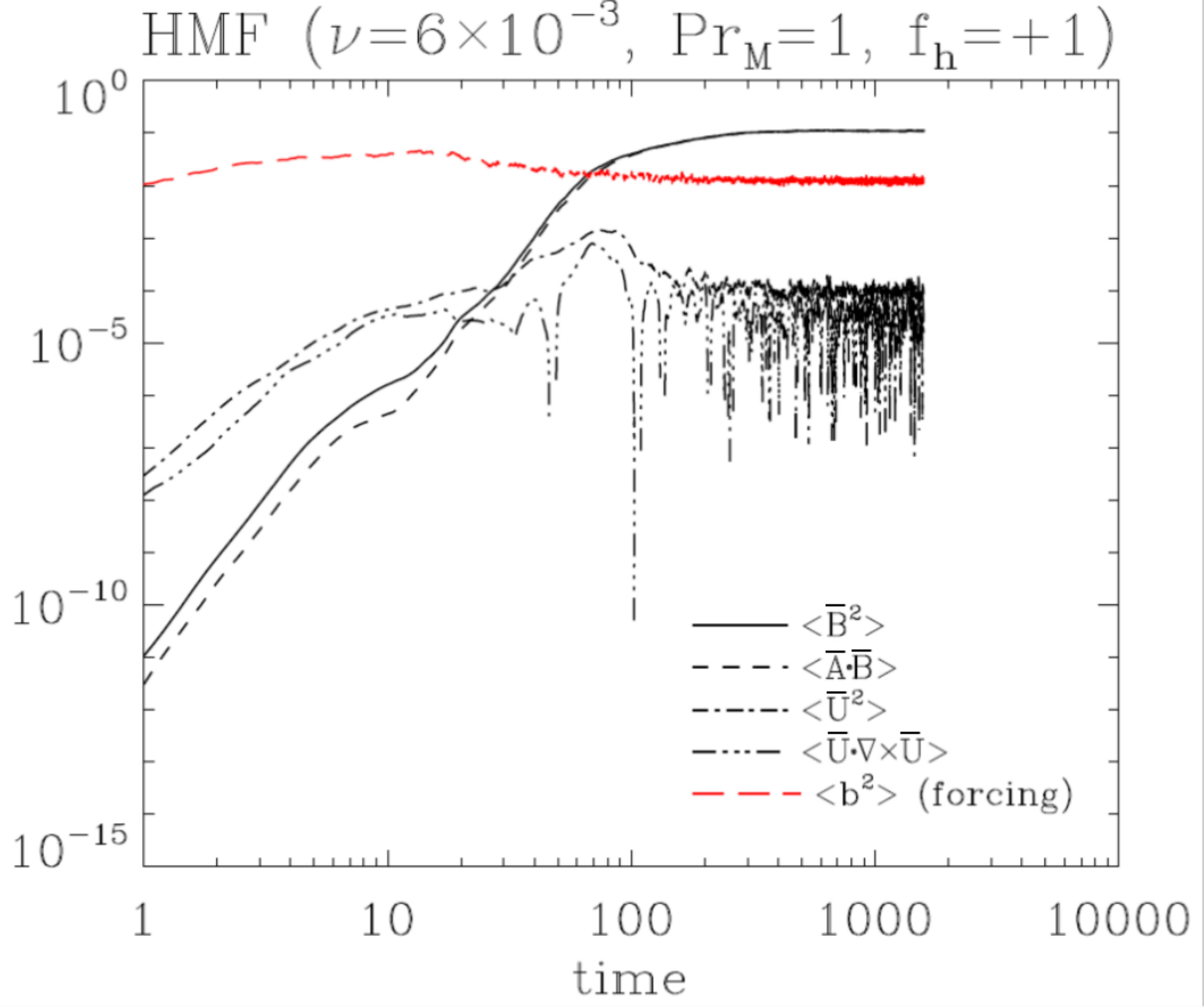}
     \label{fig1a}
}\hspace{-5 mm}
   \subfigure[]{
     \includegraphics[width=7.5 cm]{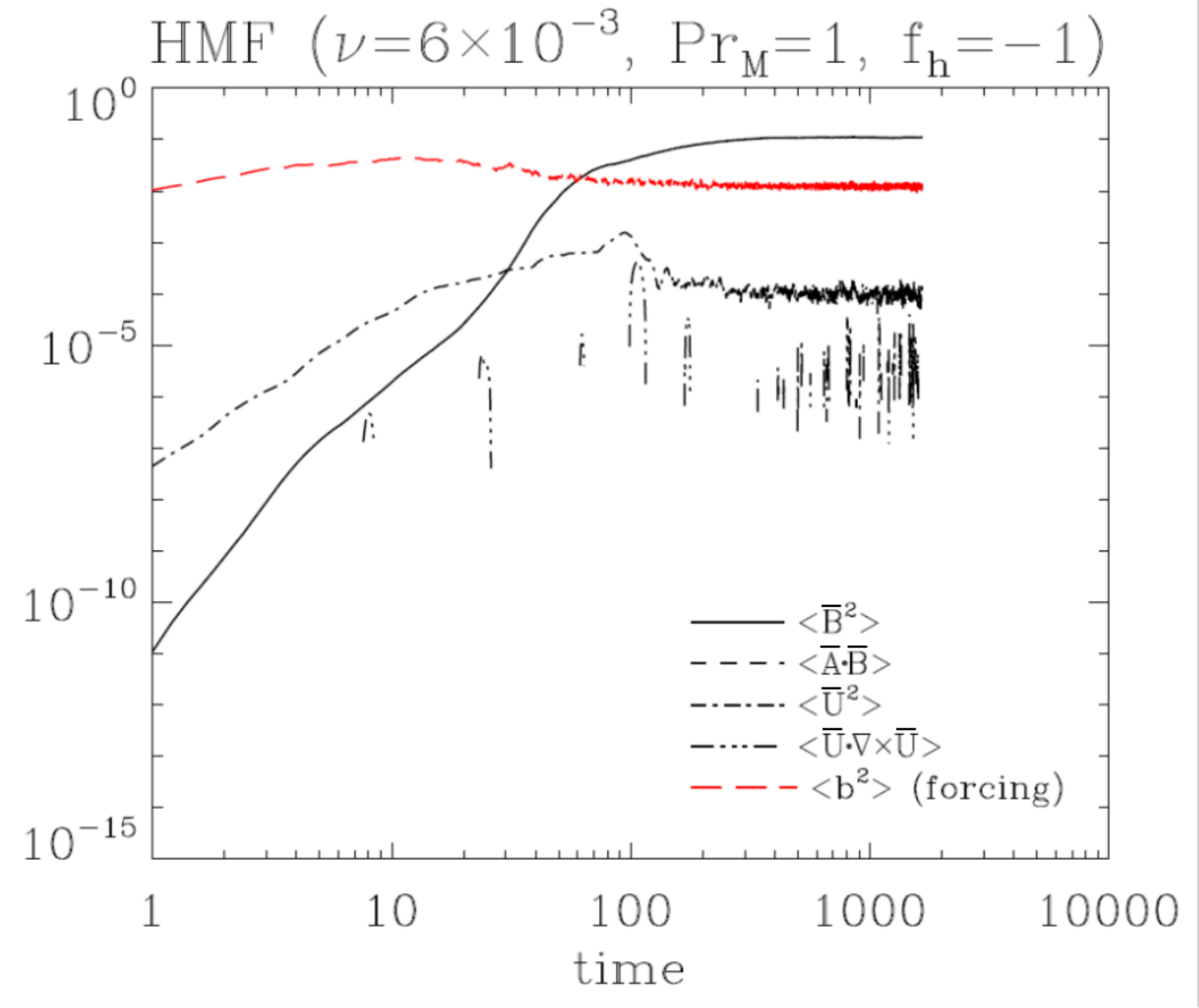}
     \label{fig1b}
   }\hspace{-5 mm}
   \subfigure[]{
     \includegraphics[width=7.5 cm]{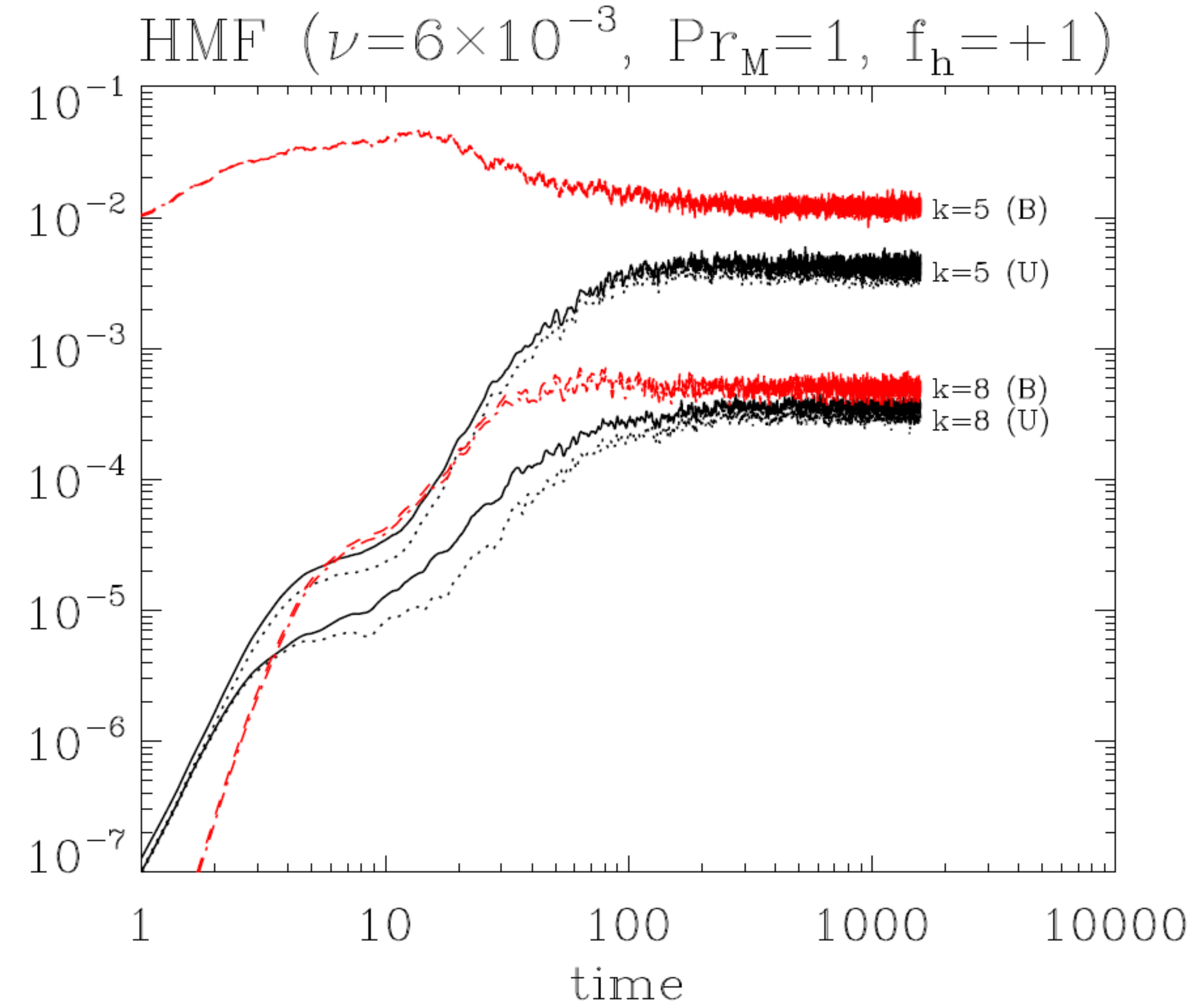}
     \label{fig1c}
   }\hspace{-5 mm}
   \subfigure[]{
     \includegraphics[width=7.5 cm]{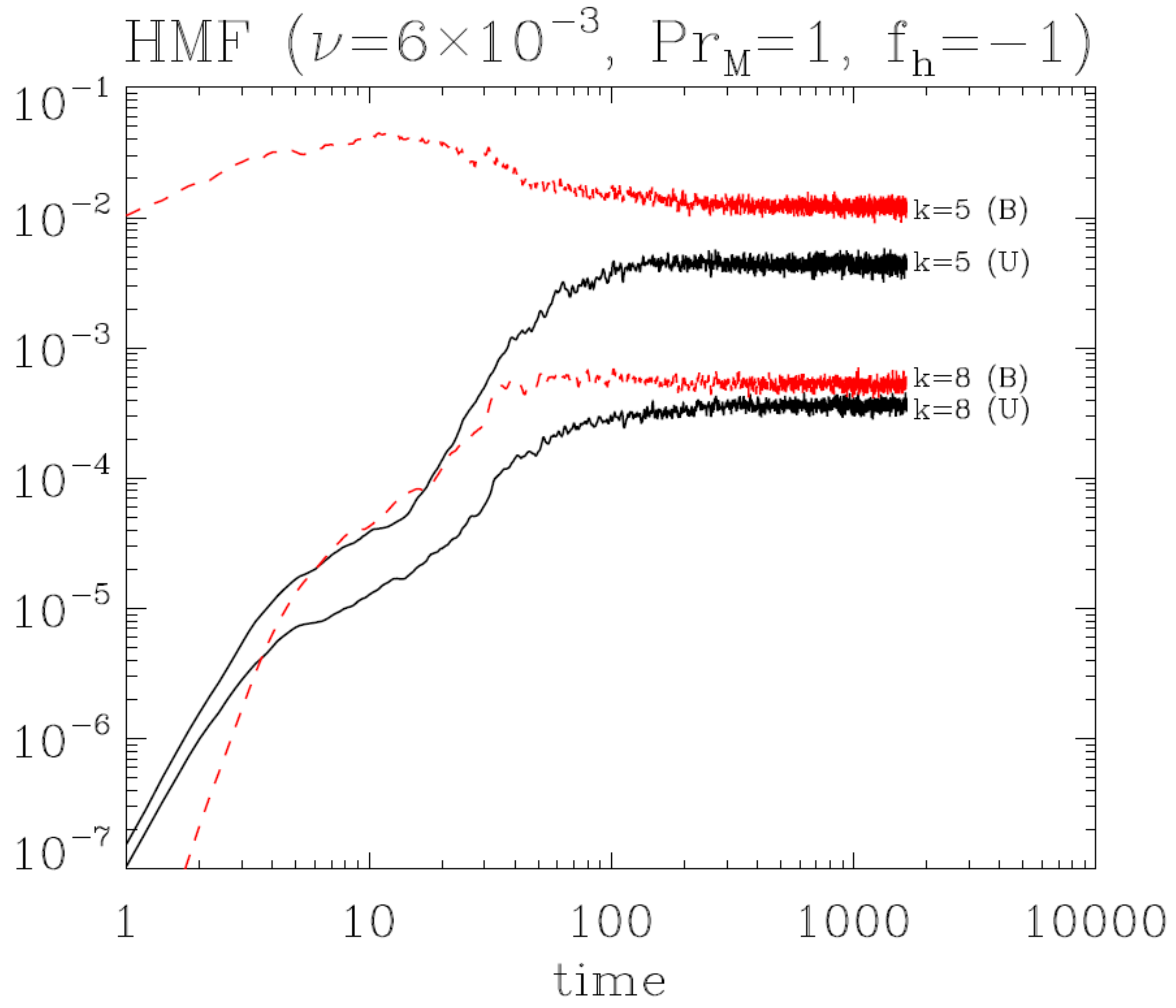}
     \label{fig1d}
   }
}
\caption{Plot (a) \& (c) show the logarithmic evolution of energy and helicity of the system forced with right handed helical magnetic field ($f_h=+1$). Plot (b) \& (d) are the same as (a) \& (c), but the system was forced with left handed helical magnetic field ($f_h=-1$). In (c), (d), $k=5$ indicates the forcing scale eddy, and $k=8$ indicates one of the small scale ones. The red dashed line means magnetic energy $\langle b^2\rangle$, and the red dotted one means its helical contribution $k\langle {\bf a}\cdot {\bf b} \rangle$. {Here, small character $u,\,b$ represent the turbulent scale regime. The symbol (B), (U) are to indicate they are for magnetic field and velocity field.} In forcing scale (k=5), magnetic energy and its helical part are practically the same so that the corresponding lines are overlapped. On the other hand, the black solid line indicates kinetic energy $\langle u^2\rangle$, and the black dotted line indicates its helical part $\langle {\bf u}\cdot \nabla \times {\bf u} \rangle/k$. In  (a) \& (c) kinetic and magnetic helicity clearly show up, but those in  (b) \& (d) are not shown except some part of large scale kinetic helicity. This indicates that the polarization of helicity in HMFD, except the large scale  velocity, is consistently decided by forcing magnetic field.}
     \label{fig1}
\end{figure*}

\begin{figure*}
\centering{
   \subfigure[]{
     \includegraphics[width=8.3 cm]{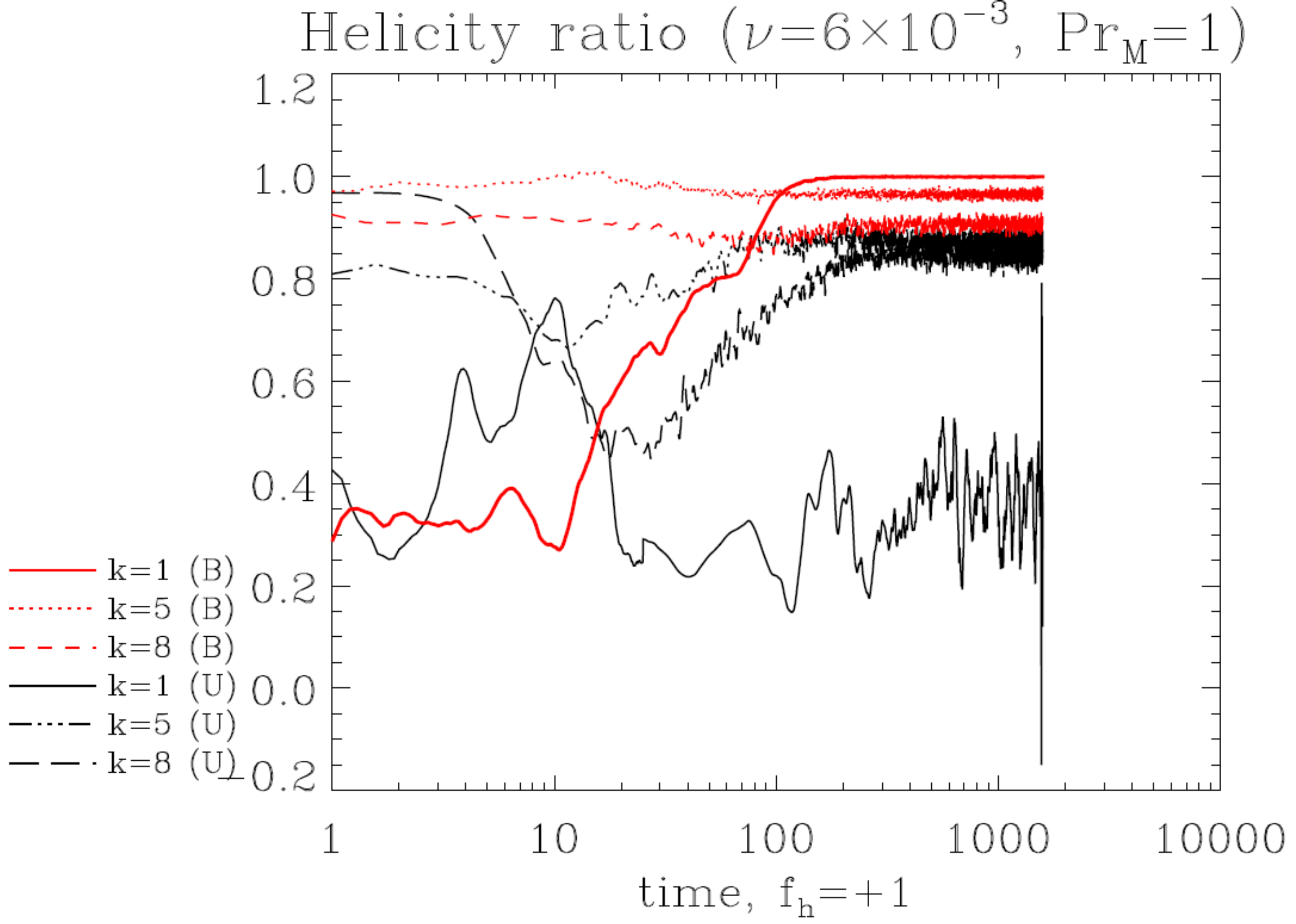}
     \label{fig2a}
}\hspace{-5 mm}
   \subfigure[]{
     \includegraphics[width=7.05 cm]{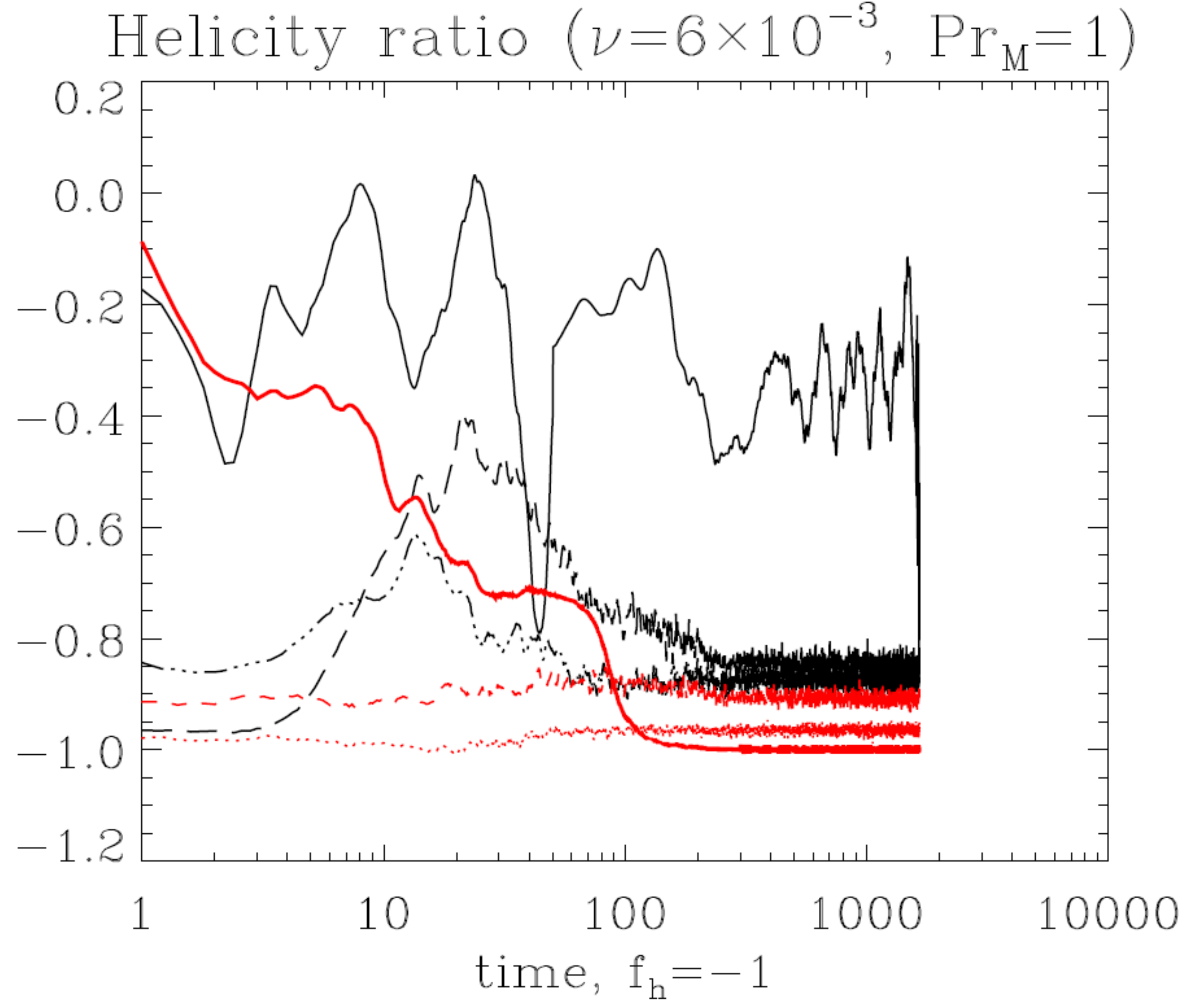}
     \label{fig2b}
   }
}
\caption{The evolution of helicity ratio: $k\langle {\bf A}\cdot {\bf B}\rangle/\langle B^2 \rangle$ for magnetic energy and helicity, $\langle{\bf U}\cdot \nabla \times {\bf U}\rangle/k\langle U^2\rangle$ for kinetic energy and helicity (k=1, 5, 8). (a) It should be noted that $f_h$ of large scale magnetic field (red thick line) is not 1 from the first. Rather, it begins from a low value and converges to 1 as the system becomes saturated. (b) $f_h$ of large scale magnetic field converges to -1. The constantly saturated magnetic helicity ratio that $k$ for the large scale field is clearly 1, not $\sqrt{2}$.  Moreover, the wavenumber does not depend on time with the normalized code data.}
\label{fig2}
\end{figure*}

\section{Numerical method}
The basic MHD equations are composed of continuity, momentum, and magnetic induction equation as follows:
\begin{eqnarray}
\frac{\partial \rho}{\partial t}&=&-{\bf U} \cdot {\bf \nabla}\rho -\rho {\bf \nabla} \cdot {\bf U},\label{continuity_equation_original}\\
\frac{\partial {\bf U}}{\partial t}&=&-{\bf U} \cdot {\bf \nabla}\mathbf{U}-{\bf \nabla} \mathrm{ln}\, \rho + \frac{1}{\rho}{\bf J}\times {\bf B}\nonumber\\&&
+\nu\big({\bf \nabla}^2 {\bf U}+\frac{1}{3}{\bf \nabla} {\bf \nabla} \cdot {\bf U}\big)+\textbf{f}_{kin},\label{momentum_equation_original}\\
\frac{\partial \mathbf{A}}{\partial t}&=& \mathbf{U}\times \mathbf{B} -\eta \nabla \times \mathbf{B}+\textbf{f}_{mag}.\label{vector_potential_induction_equation}\\
\bigg(\Rightarrow \frac{\partial \mathbf{B}}{\partial t}&=&\nabla \times (\mathbf{U}\times \mathbf{B}) +\eta \nabla^2\mathbf{B}+\nabla\times\textbf{f}_{mag}\bigg),\label{magnetic_induction_equation_original}
%\frac{\partial {\bf A}}{\partial t}&=&{\bf U}{\bf \times} {\bf B} -\eta\,{\bf \nabla}{\bf \times}{\bf B}\,\,(+\mathbf{f}_{mag})\,\,\,\,
\end{eqnarray}
Here, the symbols $\rho$, $\nu$, and $\eta$ indicate density, kinematic viscosity, and magnetic diffusivity. `{\bf U}' is in the units of sound speed $c_s$, and `{\bf B}' is normalized by $(\rho_0\,\mu_0)^{1/2}c_s$ ($\mu_0:$ magnetic permeability in vacuum. The variables in these equations are unitless.)\\

\noindent The fields $\bf U$, $\bf B$ can be separated into the large scale fields $\overline{\bf U}$, $\overline{\bf B}$ and turbulent small scale ones $\bf u$, $\bf b$. Analytically, the  evolution of $\overline{\mathbf{B}}$ can be represented as follows:
\begin{eqnarray}
\frac{\partial \overline{\mathbf{B}}}{\partial t}&\sim&\nabla \times \langle \mathbf{u}\times \mathbf{b}\rangle  +\eta \nabla^2\overline{\mathbf{B}},\label{LS_magnetic_induction_equation_raw}\\
&\sim& \nabla \times (\alpha \overline{\mathbf{B}})+(\beta+\eta)\nabla^2\overline{\mathbf{B}}
\label{LS_magnetic_induction_equation_alpha_beta}
\end{eqnarray}
{Since $\alpha$\&$\beta$ are turbulent quantities, their curl effect over the large scale is negligible. Then, with some rearrangement, we can derive Eq.~(\ref{LS_magnetic_induction_equation_alpha_beta}). And, this diffusion equation can be analytically solved with some appropriate closure theory. However, since Eq.~(\ref{continuity_equation_original})-(\ref{vector_potential_induction_equation}) should be numerically solved, we used Pencil-Code \citep{Brandenburg}(see the manual http://pencil-code.nordita.org). Pencil-code is a 6$^{th}$ order finite-difference code for compressible fluid dynamics (CFD) with magnetic field. The code solves vector potential `$\bf A$' in Eq.~(\ref{vector_potential_induction_equation}) instead of magnetic field `${\bf B}$' Eq.~(\ref{magnetic_induction_equation_original}). Solving `$\bf A$', the condition of divergence free magnetic field ($\nabla \cdot {\bf B}=0$) is met without any numerical manipulation. Moreover, magnetic helicity $H_M(=\langle {\bf A}\cdot {\bf B}\rangle)$ as well as magnetic energy $E_M(=\langle B^2/2\rangle)$ can be calculated free from {\bf the gauge issue with the assumption of the simply connected volume bounded by a magnetic surface (${\bf B}\cdot \hat{n}=0$.)}\\

\noindent {We initially gave a random seed magnetic field $B_0\sim 10^{-4}$ (no unit) to a cube simulation box $(8\pi^3$) with a periodic boundary condition. But, what determines the evolution of the system is the forcing method. We used a forcing function as follows:}
\begin{eqnarray}
{\bf f}(k,\,t)=\frac{i\mathbf{k}(t)\times (\mathbf{k}(t)\times \mathbf{\hat{e}})-\lambda |{\bf k}(t)|(\mathbf{k}(t)\times \mathbf{\hat{e}})}{k(t)^2\sqrt{1+\lambda^2}\sqrt{1-(\mathbf{k}(t)\cdot \mathbf{e})^2/k(t)^2}}.
\label{forcing amplitude fk}
\end{eqnarray}
{This is the Fourier transformed function represented by a wavenumber `$k$', helicity ratio controller $\lambda$,  and  arbitrary unit vector `$\mathbf{\hat{e}}$'. `$k$' is inversely proportional to the eddy scale $l\sim 1/k$. For instance, $k=1$ indicates the large scale regime, and $k>2$ indicates the small (turbulent) scale regime. We gave helical magnetic field (energy) at the randomly chosen wavenumber $k$, which is constrained by $\langle k\rangle_{ave}\equiv k_f \sim 5$. This forcing function can be located at Eq.~(\ref{momentum_equation_original}) (KFD), or Eq.~(\ref{vector_potential_induction_equation}) (MFD). And, if $\lambda$ is $+(-)1$, the forcing energy is fully right (left) handed helical field $i\mathbf{k}\times \mathbf{f}=\pm k\mathbf{f}$. But, $\lambda=0$ yields a nonhelical forcing source.}\\

{\bf \noindent Of course, the forcing source in nature is different from Eq.(\ref{forcing amplitude fk}). In Appendix, we introduced two examples of magnetic forcing dynamo $f_{mag}$: Biermann battery effect and neutrino-lepton interaction. One of the most essential differences between them is magnetic helicity that decides the direction of magnetic energy transfer. Biermann effect does not have helicity so that the magnetic energy cascades toward the smaller scale regimes (small scale dynamo). In contrast, neutrino-lepton interaction yields magnetic helicity, i.e., $\alpha$ effect, which can play an important role in large scale dynamo.\\}

%\noindent MFD is likely to occur in the magnetically dominated systems such as the astrophysical corona, disk, and experimental plasma lab \cite{Tzeferacos}. It should be noted that dynamo is an electromagnetic phenomenon. Magnetic field is generated by current density according to Biot-Savart law and Maxwell equation. The propagation of magnetic field with the massive plasma particles is also essentially the induction of magnetic field, not the stretching or folding of the field as MHD equations indicate.\\

%\noindent On the other hand, the external forcing source `$\bf f$' in Eq.~(\ref{LS_magnetic_induction_equation_alpha_beta}) does not explicitly appear. So, it may be thought that there is no forcing effect. However, `$\bf f$', given to the small scale regime ($k_f\,>2)$, is included in $\alpha$ and $\beta$.

%Except some special cases that emphasize the effect of `$\bf f$' (Eq.~32 in \cite{Park2014}), $\alpha$\&$\beta$ terms are considered to include `$\bf f$'. At present, their exact formal definitions are not yet known. We calculate them using large scale magnetic energy and magnetic helicity without artificial manipulation in HMFD.\\

\noindent We prepared for the two systems with a unit magnetic Prandtl number $Pr_M=\eta/\nu=1$. $\eta=\nu=6\times 10^{-3}$, and numerical resolution is $400^3$. They were forced by Eq.~(\ref{forcing amplitude fk}) with fully helical magnetic field ($\lambda=+1$ or $-1$ at $k=5$). We used basic data set such as kinetic energy $E_V(=\langle U^2\rangle/2)$, magnetic energy $E_M(=\langle B^2\rangle/2)$, kinetic helicity $\langle {\bf U}\cdot \nabla\times {\bf U}\rangle$, and magnetic helicity $\langle {\bf A}\cdot {\bf B}\rangle$. The stability of code and data have been verified.

%As the last term implies, the neutrino interaction generates helical magnetic energy so that the forcing function can be replaced by $\sim \alpha_{\nu}{\bf B}$.

\section{Numerical Result}
%Fig.~1 shows how the magnetic field and velocity field evolve in the magnetically driven system.
The system in Fig.~1(a), 1(c) is forced by the fully positive (right-handed) helical magnetic field (red dashed line, helicity ratio of forcing energy: $f_h\equiv k_f\langle {\bf a}\cdot {\bf b}\rangle/\langle b^2\rangle=1,\,k_f=5$ forcing wavenumber). In contrast, the system in the right panel Fig.~1(b), 1(d) is forced by the fully negative (left-handed) magnetic helicity ($f_h=-1$). Peak speed $U$ is $\sim 2\times 10^{-3}$, and magnetic Reynolds number is defined as $Re_M\equiv UL/\eta\sim 2\pi/3$, where $L$, $\eta$ are $2\pi$ and $6\times 10^{-3}$, respectively. In HMFD, the least amount of magnetic energy is transferred to plasma.\\

\noindent In Fig.~1(a), {large scale magnetic energy $\langle {\overline B}^2\rangle$ (=$2\overline{E}_M$, $k=1$, solid line)} grows to be saturated at $t\sim 100$. {Along with $\langle {\overline B}^2\rangle$, the large scale magnetic helicity $\langle {\overline {\bf A}}\cdot {\overline {\bf B}}\rangle$(dashed line) evolves keeping the relation of $\langle {\overline B}^2\rangle\ge k\langle{\overline {\bf A}}\cdot {\overline {\bf B}}\rangle$($k=1$). Also, kinetic energy $\langle {\overline U}^2\rangle$ in the large scale grows keeping $\langle {\overline U}^2\rangle \geq \langle {\overline {\bf U}}\cdot \nabla \times {\overline {\bf U}}\rangle/k$.} But the direction of kinetic helicity fluctuates from positive to negative as the discontinuous cusp line implies in this log-scaled plot. {Similarly, Fig.~1(c) shows that evolving small scale magnetic energy $\langle b^2 \rangle$ and kinetic energy $\langle u^2 \rangle$ with their helical part $\langle {{\bf u}}\cdot \nabla \times {{\bf u}}\rangle/k$ and $k\langle {{\bf a}}\cdot {{\bf b}}\rangle$ ($k\geq 2$).} The fields grow and get saturated like the large scale field, but the saturation occurs earlier than that of the large scale field because of their smaller eddy turnover time.\\

\noindent Fig.~1(b), 1(d) show the growth of kinetic and magnetic energy in the system forced by the fully left-handed magnetic energy. Basically, they evolve consistently in comparison with Fig.~1(a), 1(c). But, the kinetic helicity and magnetic helicity are invisible in this logarithmic plot indicating their left-handed (negative) chirality. The direction of  helicity in the magnetically forced system tends to be consistent with that of forcing energy. This is the opposite tendency of HKFD (helical kinetic forcing dynamo). Nonetheless, the practically same growth of energy shows that the chirality of the forcing energy is not a determinant to the evolution of the plasma system.\\

\noindent Fig.~2 shows the evolving magnetic helicity ratio $f_h\equiv k\langle  {\bf a}\cdot {\bf b}\rangle/\langle b^2\rangle$ and kinetic helicity ratio $\langle  {\bf u}\cdot \nabla \times {\bf u}\rangle/k\langle u^2\rangle$ for $k$=1, 5, 8. Left and right panel are for the right handed case ($f_h=1$) and left handed one ($f_h=-1$), respectively. The helicity ratio of large scale ${\overline B}$ is eventually saturated at $f_h=+1$ ($-1$), and that of the small scale u \& b reaches to the value less (larger) than `$1$ ($-1$)'. However, the helicity ratio of ${\overline U}$ is as low as $\sim 0.25$ ($-0.25$). The magnetic helicity ratio less than `1' in the small scale regime shows that the small scale magnetic field substantially accelerates the large scale plasma motion. {Plasma in the large scale regime is driven by Lorentz force ${\bf J}(p)\times {\bf B}(q)$, where the wave numbers are constrained by $p+q=1$. This implies that the eddies associated with $p$, $q$ are very close to each other in the small scale regime and nearly out of phase. Their helicity ratios smaller than $1$ implies Lorentz force forcing the large scale eddy meaningfully grows.} But, the effect of the large-scale magnetic field on the plasma is limited: $\overline{\bf J}\times \overline{\bf B}\sim 0$. The saturated helicity ratio `$f_h=$1' for the large scale field indicates that $k$ for $\overline{\bf B}$ is  definitely `1'. Also, the initial largest energy level with `$k=5$' implies that the scale with this wavenumber is forced by an external energy source, which is consistent with our code setting.\\

\noindent Fig.~3 includes the temporally evolving $\alpha$ \& $\beta$ effect and the large scale magnetic energy $2\overline{E}_M$. Left (right) panel shows the evolution of $E_M$, $\alpha$ and $\beta$ effect for $f_h= 1$ ($f_h=-1$). $\alpha$ effect for $f_h= 1$ positively oscillates and {\bf decreases significantly} as $E_M$ gets saturated. In contrast, $\alpha$ effect for $f_h=-1$ negatively oscillates before it disappears.  $\alpha$ effect is quenched much earlier than the slowly evolving $E_M$. The decreasing oscillation in both cases implies that $\alpha$ effect does not play a decisive role in the growth of the large scale magnetic field. Conversely, $\beta$ retains the negative value in both cases and has a much larger size than that of $\alpha$. This negative $\beta$, combined with the negative Laplacian $\nabla^2\rightarrow-k^2$ in Fourier space, can be considered as the actual source of the large scale magnetic field. This is contradictory to the current dynamo theory concluding that $\beta$ is always positive to diffuse magnetic energy. We will show that this conventional inference is valid only for the ideally isotropic system with reflection symmetry. When the symmetry is broken, `$\bf u$' in the small scale regime can yield the anti-diffusing effect of magnetic field.\\

\noindent In Fig.~4, we compared $\nabla\times \langle {\bf u}\times {\bf b}\rangle$ (black solid line) with $\nabla \times (\alpha \overline{\bf B}-\beta\nabla\times \overline{\bf B})$ (red dashed line) to verify Eq.~(\ref{LS_magnetic_induction_equation_alpha_beta}) and Eq.~(\ref{alphaSolution3}), (\ref{betaSolution3}). {The former uses only the simulation data for the large scale magnetic energy $\overline{E}_M$, but the latter requires the data of large scale magnetic helicity $\overline{H}_M$ in addition to $\overline{E}_M$}. {Although they use different types of data and formulas,} they are coincident in the transient mode ($t<\sim 100$ and in the range of $10^{-8}-10^{-2}$). {Note that Eq.~(\ref{alphaSolution3}), (\ref{betaSolution3}) are valid until the system gets saturated where  $\overline{H}_M\sim 2\overline{E}_M$. As Fig.~1, 2 show, $\overline{H}_M$ is different from $2\overline{E}_M$ in the transient state}. As the field becomes saturated, $\overline{E}_M$ and $\overline{H}_M$ are so close that the logarithmic function diverges. For $f_h=-1$, we used absolute values for a clear comparison.\\

\noindent Fig.~5, 6 show field structure models. They are introduced to explain the dynamo process in an intuitive way. We will discuss the mechanism in detail soon.\\

\noindent Fig.~7 is for the typical kinetic small scale dynamo. Nonhelical random velocity field was driven at $k=5$. The plot includes large scale kinetic energy $E_V$ and magnetic energy $E_M$. $Re_M$ is approximately $80$. In comparison with LSD, $E_M$ grows a little bit, and $E_V$ is not quenched. Most magnetic energy is transferred to the small scale regime, and its peak is located at $k\sim 10$. These plots are to compare Fig.~1(a), 1(b).\\

\noindent Fig.~\ref{fig7a}, \ref{fig7b} in appendix include the simulations of solar magnetic field using Eq.~(\ref{Solar_poloidal_magnetic_field}) and (\ref{Solar_toroidal_magnetic_field}). The horizontal axis means `scaled time' ($0.01s\rightarrow 15.53$ years), and the vertical line indicates the `latitude' ($0-\pi/2$: northern hemisphere, $\pi/2-\pi$: southern hemisphere). The color indicates the phase of a net magnetic field (toroidal ${\bf B}_{tor}$ + poloidal ${\bf B}_{pol}$). The simulation in the left panel is the reproduction of \cite{Jouve} without Babcock effect. It shows the period of 16.31 years for the one complete cycle of solar magnetic field: amplification-annihilation-reverse. On the contrary, in right panel, the tidal effect of planets on the Solar tachocline is added to $\alpha$. The period elevates up to 21.74 years without manipulating the numerical variables. These plots show the practical applicability of Eq.~(\ref{magnetic_induction_equation_original}).\\

\noindent Fig.~\ref{fig8} in appendix shows the time evolving $1/3 \langle u^2\rangle$ and $1/6 \langle {\bf u}\cdot \nabla \times {\bf u}\rangle$. This plot is to make sure the overall $\beta$ effect in HMFD is negative. We will also briefly introduce Kraichnan's result \citep{Kraichnan}. He expanded $\alpha$ with the unknown time constant $\tau_2$ and correlation factor $A(x-x')$, and they work as a negative magnetic diffusivity.\\

\noindent Fig.~\ref{fig9} in appendix shows the effect of helical velocity field in magnetic diffusivity. Fully helical kinetic forcing is turned off at $t\sim 200$ followed by nonhelical kinetic forcing where $\langle {\bf u}\cdot {\nabla \times \bf u}\rangle=0$. After $t\sim 200$, turbulent magnetic diffusivity becomes positive, which diffuses magnetic field in the system. The growth of large scale magnetic field is suppressed.

\begin{figure*}
\centering{
   \subfigure[]{
     \includegraphics[width=7.5 cm]{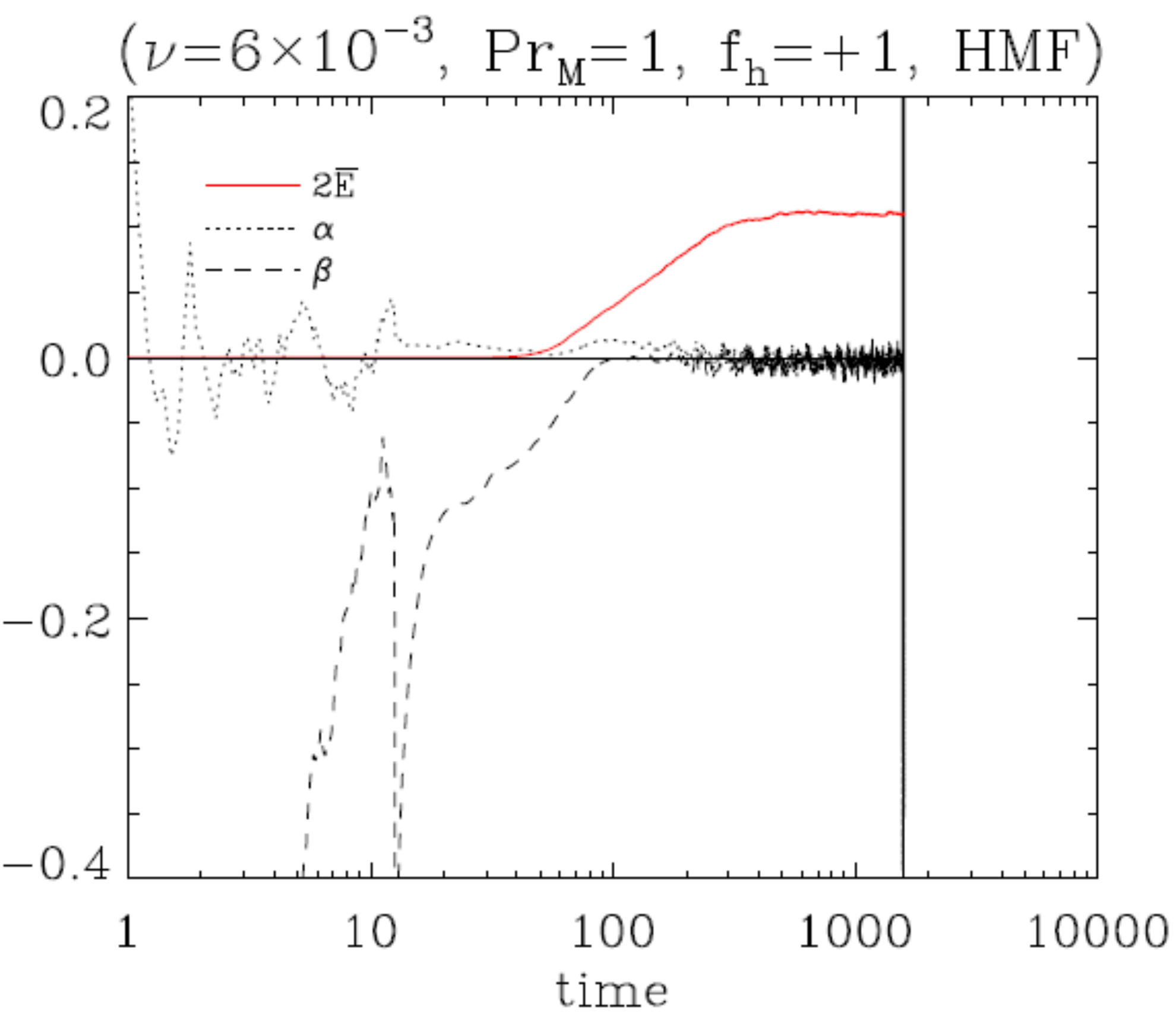}
     \label{fig3a}
}\hspace{-5 mm}
   \subfigure[]{
     \includegraphics[width=7.5 cm]{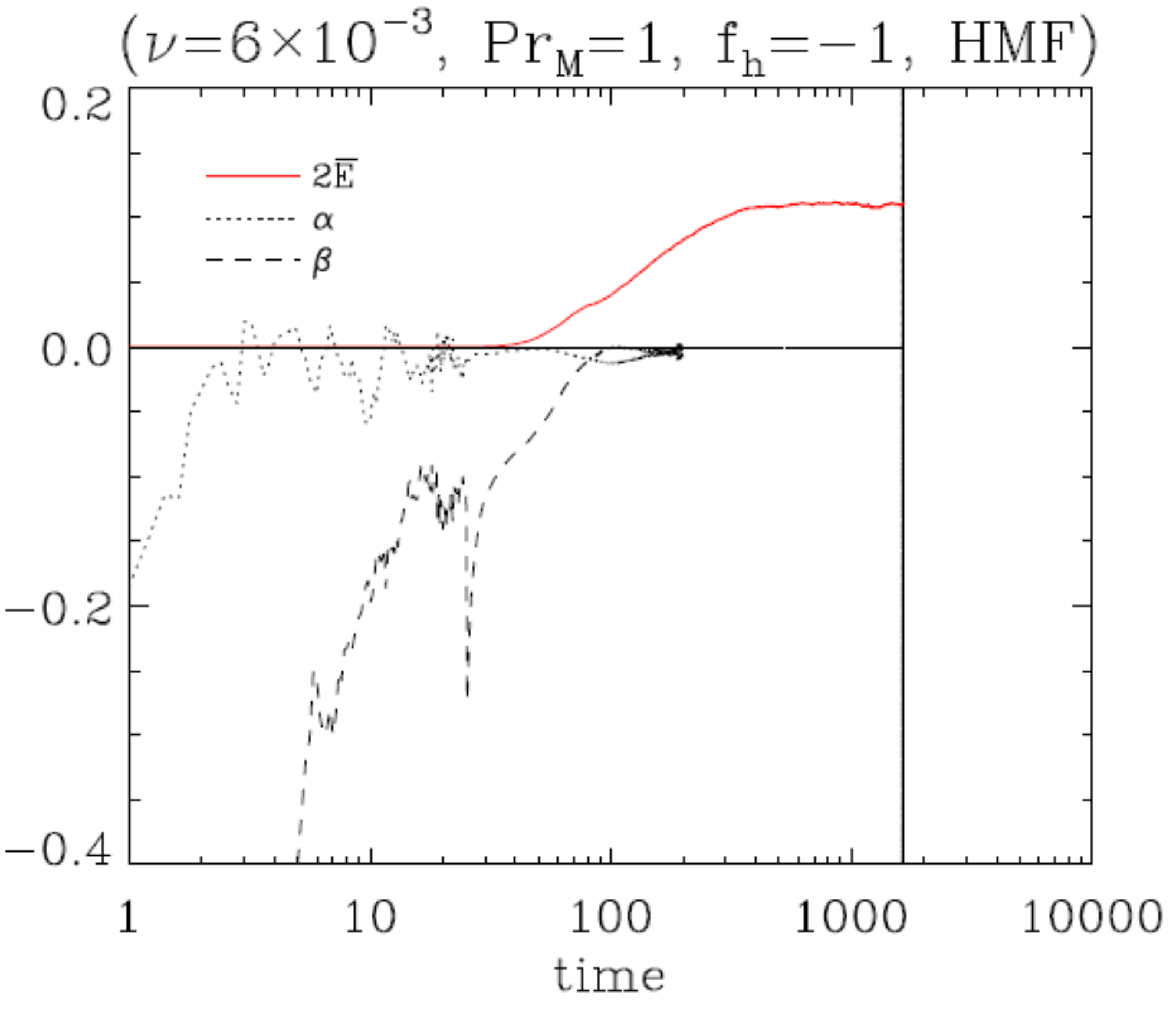}
     \label{fig3b}
   }
}
\caption{$\alpha(t)$, $\beta(t)$, and $\langle {\overline B}^2\rangle$ for $f_h=1$ and $-1$. The small and early quenching $\alpha$ effect shows its limited effect on the growth of 2${\overline E}_M$.}
\end{figure*}

\begin{figure*}
\centering{
   \subfigure[]{
     \includegraphics[width=7.5 cm]{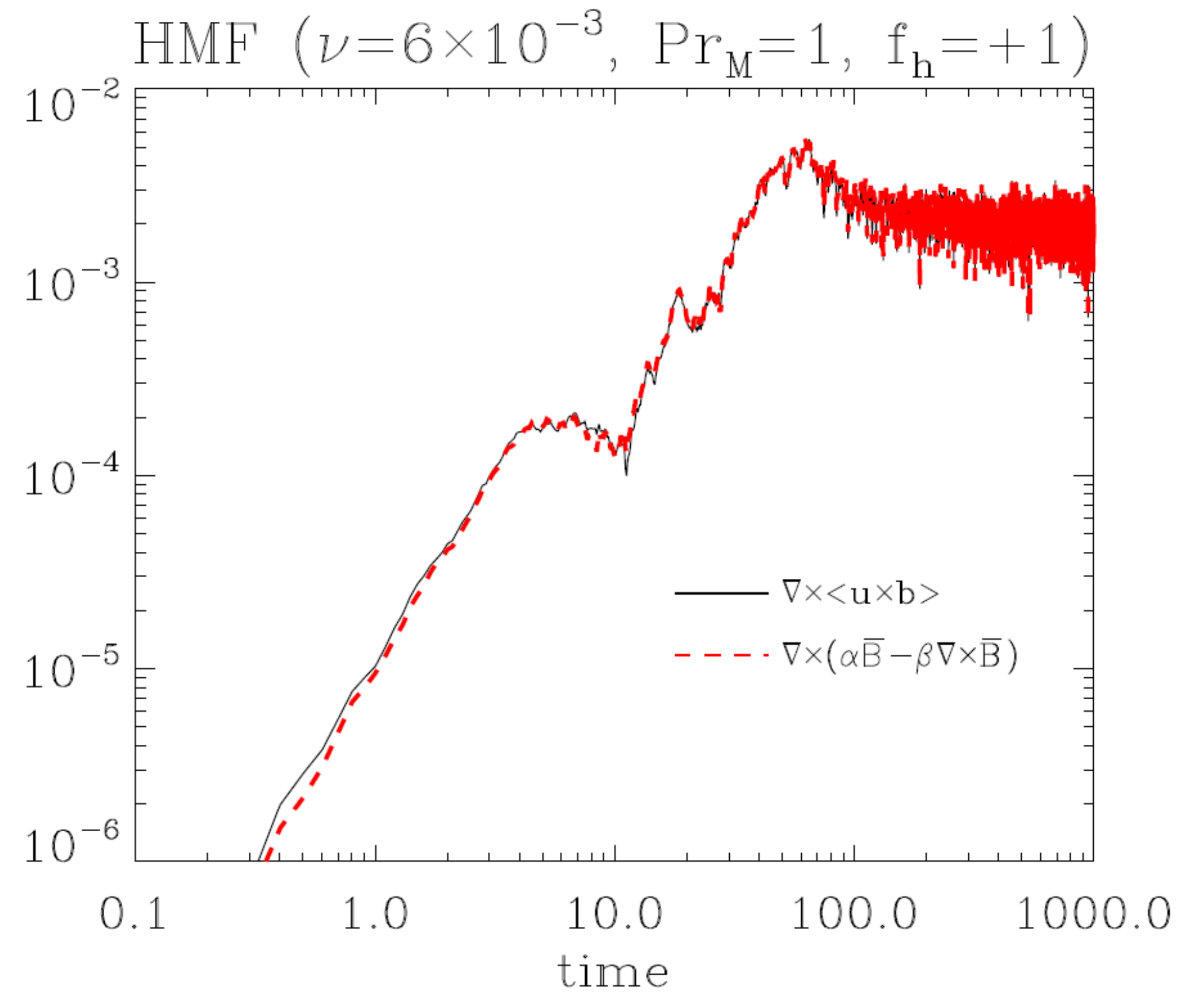}
     \label{fig3a1}
}\hspace{-5 mm}
   \subfigure[]{
     \includegraphics[width=7.5 cm]{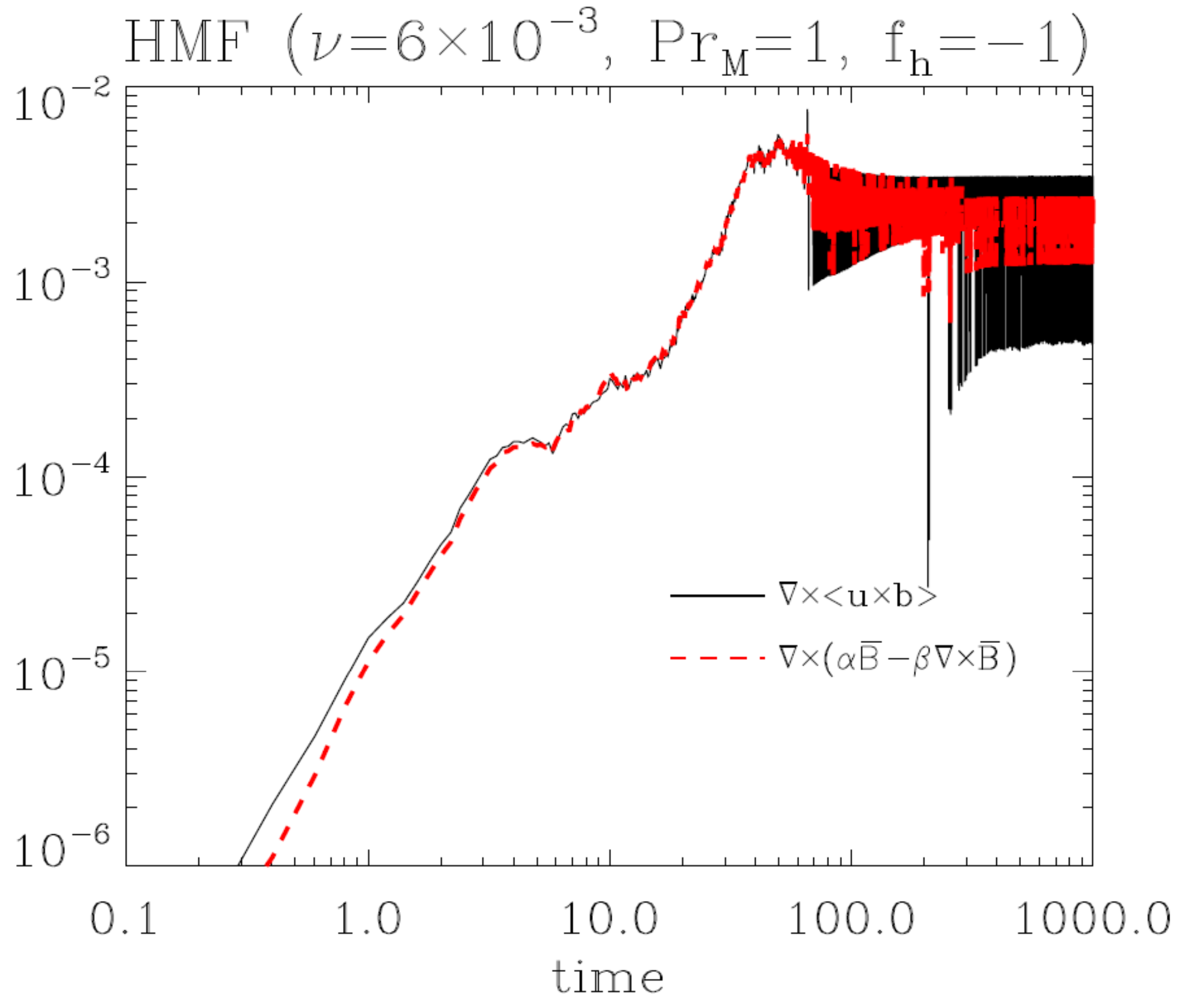}
     \label{fig3b1}
   }
}
\caption{Comparison of EMF and $\alpha$ \& $\beta$ approximation.  These plots are to verify the separation of $\alpha$\&$\beta$ from $EMF$. They are from different physical quantities and formulas but yield quite coincident results. $\partial {\overline{\bf B}}/\partial t -\eta \nabla^2{\overline{\bf B}}$ for $\nabla\times\langle {\bf u}\times {\bf b}\rangle$ is the explicit function of $\overline{E}_M$, $\eta$, and $t$. But, $\alpha$\&$\beta$ are functions of $\overline{E}_M$ and $\overline{H}_M$ and $\eta$. Average is taken for the clear comparison.}
\label{fig3}
\end{figure*}

\begin{figure*}
\centering{
   \subfigure[Helical Magnetic forcing]{
     \includegraphics[width=6.7 cm]{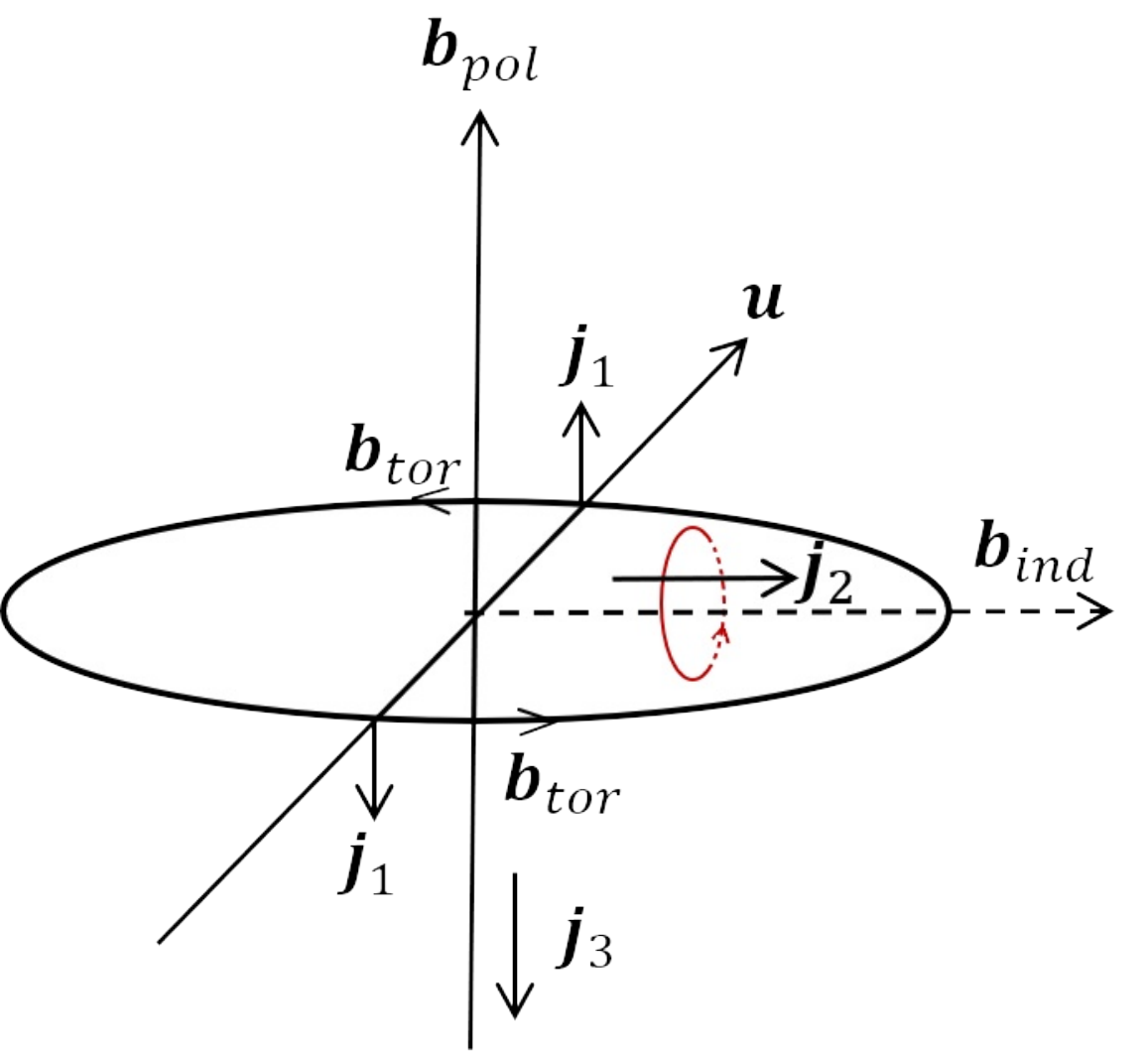}
     \label{fig4a}
}\hspace{0 mm}
   \subfigure[Induced helical kinetic forcing]{
     \includegraphics[width=7.7 cm]{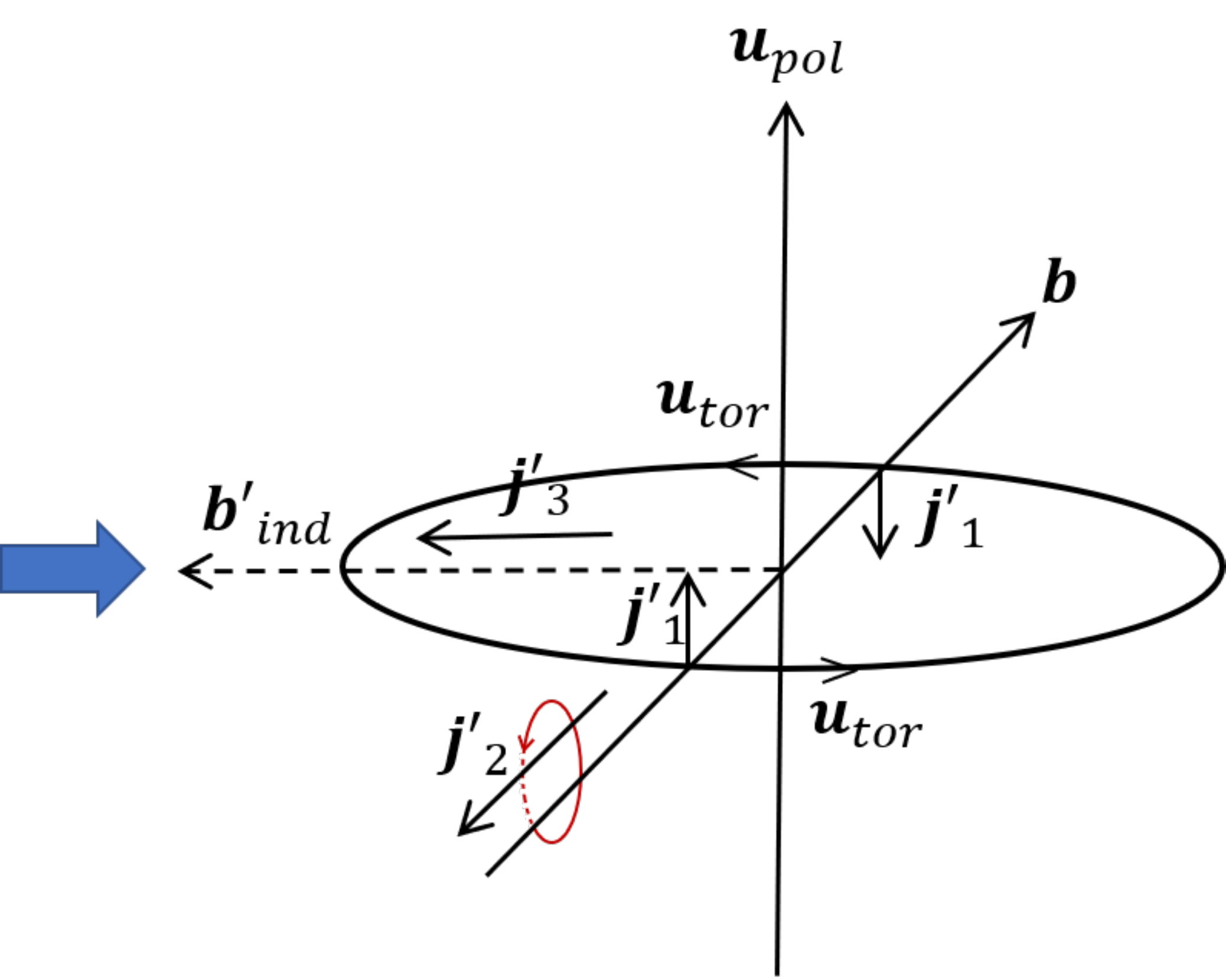}
     \label{fig4b}
   }
}
\caption{{${\bf j}_{1,\,up}$ and ${\bf j}_{1,\,down}$ are represented as ${\bf j}_{1}$ for simplicity. (a) Right handed$(+)$ magnetic helicity yields $+|{\bf j}_{2}\cdot {\bf b}_{ind}|$, $-|{\bf j}_{3}\cdot {\bf b}_{pol}|$, and right handed kinetic helicity. (b) Induced kinetic helicity also generates $-|{\bf j}'_{2}\cdot {\bf b}|$, $+|{\bf j}'_{3}\cdot {\bf b}'_{ind}|$ with $\bf b$.}}
\end{figure*}

\begin{figure*}
\centering
   {
     \includegraphics[width=14 cm]{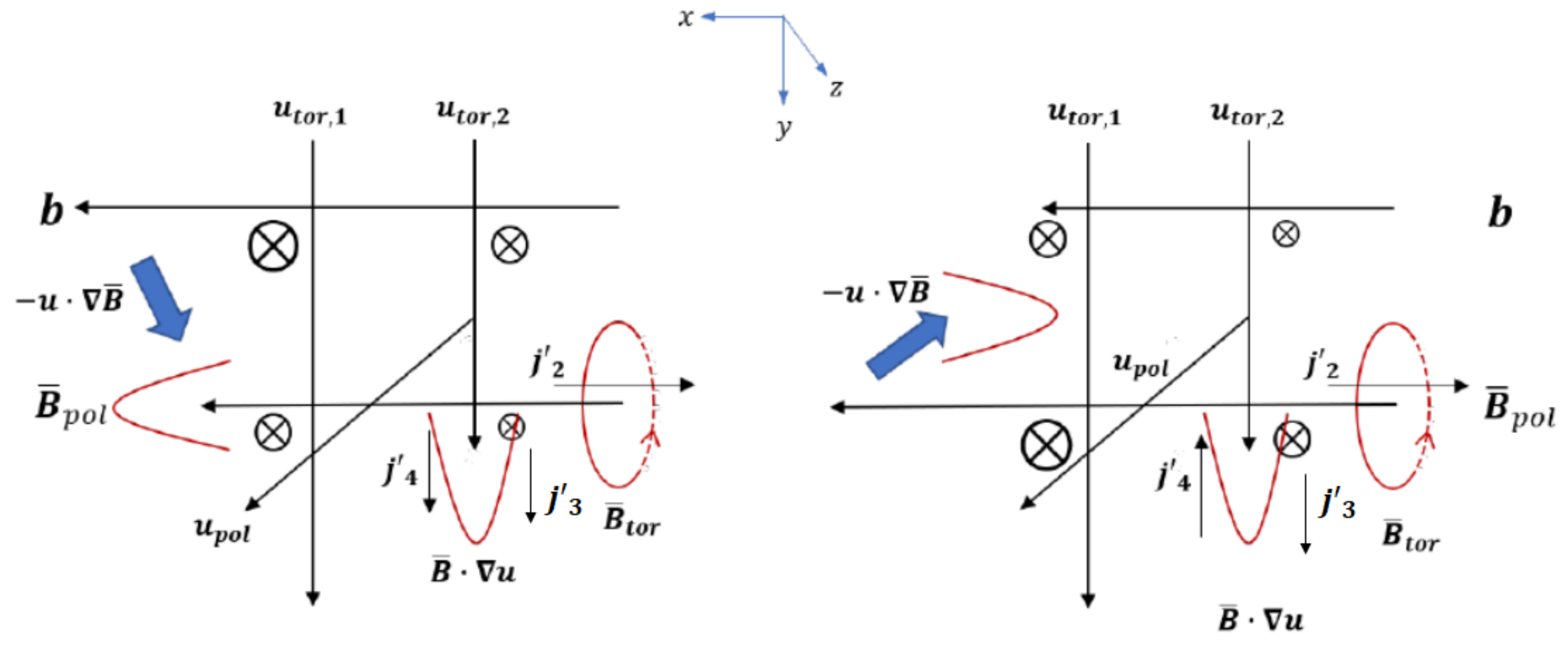}
     \label{fig5}
}
\caption{More detailed field structure based on $EMF$: $\nabla \times ({\bf U}\times {\bf B})=-{\bf U}\cdot \nabla {{\bf B}}+{\bf B}\cdot \nabla {{\bf U}}$. These structures correspond to Fig.~5(b). The left field structure is for the early time regime while ${\overline B}<b$. Right structure is for the magnetic back reaction with ${\overline B}\gtrsim b$ at later time regime, which is negligible in the magnetically forced system. The symbol `$\bigotimes$' means the direction ($-\hat {z}$) of induced current density ${\bf J}\sim {\bf U}\times {\bf B}$, and its size indicates the relative strength. {In the figure, $\bf u$ and $\bf b$ indicate turbulent velocity and magnetic field. We use $\bf \overline{\bf B}$ for the large scale magnetic field through the paper. But, in this plot, it is used for the relatively larger scale magnetic field compared to $\bf b$. Also, the subindex `$pol$' is added to $\overline{\bf B}$ to discriminate $\overline{\bf B}_{tor}$.}}
\end{figure*}

\section{Discussions on Theoretical analysis}
We have shown that magnetic energy can be inversely cascaded in the system forced by the helical magnetic field ($\nabla \times {\bf B}=\lambda{\bf B}$). {HMFD has some unique features compared to HKFD.} Helical magnetic field, which makes Lorentz force $zero$, exerts a force on the system leading to the generation of helical velocity field with the same chirality. Moreover, the conservation of magnetic helicity is not valid anymore. We study the internal interaction of $\bf U$ \& $\bf B$ using field structure model and analytic method beyond conventional theory and phenomenological rope dynamo model. {Although MHD theory is based on hydrodynamics}, the generation and transport of $B$ field are innately electromagnetic phenomena constrained by the plasma motion. Note that plasma kinetic energy is converted into magnetic energy only through EMF ${\bf U}\times {\bf B}\sim \eta{\bf J}\sim \bf E$, which is different from mechanical force. {This electromagnetic force drives the charged plasma particles to produce current density which is the source of magnetic field.}

\subsection{Field structure model}
\subsubsection{$\alpha$ effect}
The right handed magnetic structure in Fig.~5(a) is composed of the toroidal magnetic component  ${\bf b}_{tor}$ and poloidal part ${\bf b}_{pol}$. Statistically, ${\bf b}_{tor}$ and ${\bf b}_{pol}$ are not distinguished in the homogeneous and isotropic system. But, if we remove reflection symmetry from the system, ${\bf b}_{tor}$ and ${\bf b}_{pol}$ become independent components playing different roles with ${\bf u}$.\\

\noindent The interaction between $\bf u$ and ${\bf b}_{tor}$ yields current density, i.e., ${\bf u}\times{\bf b}_{tor}\sim{\bf j}_{1,\,down}$ and ${\bf j}_{1,\,up}$  in the front and back. These two components induce a new magnetic field ${\bf b}_{ind}$. At the same time, ${\bf u} \times {\bf b}_{pol}$ generates ${\bf j}_2$. This current density forms the right handed magnetic helicity with ${\bf b}_{ind}$: $\langle {\bf j}_2\cdot {\bf b}_{ind} \rangle\rightarrow k_2^2\langle {\bf a}_2\cdot {\bf b}_{ind} \rangle$, which is a (pseudo) scalar to be added to the system.  There is also a possibility that $\bf u$ and ${\bf b}_{ind}$ induce ${\bf j}_3$ yielding the left handed magnetic helicity $\langle {\bf j}_3\cdot {\bf b}_{pol} \rangle$. However, the induced magnetic field from ${\bf j}_3$ is weakened by the externally provided ${\bf b}_{tor}$.\\

\noindent On the other hand, ${\bf j}_1\times {\bf b}_{tor}$ and ${\bf j}_2 \times {\bf b}_{pol}$ generate Lorentz force toward $-\bf u$, which may look just suppressing plasma motion. However, ${\bf j}_2\times {\bf b}_{tor}$ at the right and left end yields an rotation effect, which is toward `$-\bf u$'. This rotation with those two interactions generates the right handed kinetic helicity $\langle {\bf u}\cdot \nabla \times {\bf u}\rangle$. The interaction between the current density and magnetic field makes two effects. Magnetic pressure effect $-\nabla B^2/2$ from ${\bf j}_1\times {\bf b}_{tor}$ and ${\bf j}_2 \times {\bf b}_{pol}$ suppresses the plasma motion with thermal pressure $-\nabla P$. And, ${\bf j}_2 \times {\bf b}_{tor}$ creates a rotational force to form kinetic helicity with the two suppressing effects. As Fourier transformed Lorentz force ${\bf j}({\bf p}) \times {\bf b} ({\bf q})\sim \partial {\bf u}({\bf k})/\partial t$ shows, the wavenumbers `${\bf p}$' and `${\bf q}$' are constrained by the relation of ${\bf p}+{\bf q}={\bf k}$.\\

\noindent {The induced right handed kinetic helicity in Fig.~\ref{fig4b} again generates ${\bf j'}_1$ in the front and back. The spatially inhomogeneous current density from ${\bf j'}_{1,\,up}$ and ${\bf j'}_{1,\,down}$ induces ${\bf b}'_{ind}$ leading to ${\bf j}'_2$ with ${\bf u}_{pol}$, i.e., ${{\bf j}'_2\sim {\bf u}_{pol}\times {\bf b}'_{bind}}$}. Then, ${\bf j}'_2$ forms the left handed magnetic helicity with ${\bf b}$. Also, ${\bf u}_{pol}\times {\bf b}$ yields ${\bf j}'_3$ leading to right handed magnetic helicity. If all interactions are summed up, magnetic helicity in the system is $+|\langle {\bf j}_2\cdot {\bf b}_{ind}\rangle|$$-|\langle {\bf j}_3\cdot {\bf b}_{pol}\rangle|$$-|\langle {\bf j}'_2\cdot {\bf b}\rangle|$$+|\langle {\bf j}'_3\cdot {\bf b}'_{ind}\rangle|$ qualitatively. Comparing this result with $\alpha$ quenching in Fig.~\ref{fig3a}, we may question what indeed amplifies magnetic field and determines the net magnetic helicity. There is one more term to be considered, $\beta$ effect.

\subsubsection{$\beta$ effect}
Fig.~6 is the more detailed right handed helical kinetic structure of Fig.~\ref{fig4b}. It is based on the geometrical meaning of `$\nabla \times ({\bf u}\times {\overline{\bf B}})\sim {\overline{\bf B}}\cdot \nabla {\bf u}-{\bf u}\cdot \nabla {\overline{\bf B}}$'. Here, we name `$-{\bf u}\cdot \nabla {\overline{\bf B}}$' as `local transfer (advective) term', and we call `${\overline{\bf B}}\cdot \nabla {\bf u}$'  `nonlocal transfer term'. The symbol `$\otimes$' means the direction of current density `${\bf J}_i$' heading for $-\hat{\bf z}$.
{It is from ${\bf u}_{tor,\,i}\times {\bf b}(or\, \overline{\bf B})$. And, the size of $\otimes$ implies its relative strength.} Its distribution is spatially inhomogeneous so that the nontrivial curl effect generates the magnetic fields toward ${\bf \hat{x}}$ (locally transferred $\int d\tau (-{\bf u}\cdot \nabla {\overline{\bf B}})$) and $\bf {\hat y}$ (nonlocally transferred $\int d\tau ({\overline{\bf B}}\cdot \nabla {\bf u})$). Their net magnetic field ${\bf b}_{net}$ becomes a new seed field for the next dynamo process. As the net magnetic field grows, it approaches to the velocity field `$\bf u$' so that `${\bf u}\times {{\bf b}_{net}}$' itself decreases. The field gets saturated eventually if there is no additional reason e.g., frozen field or helicity.\\

%This nonlocal transfer term actually corresponds to ${\bf b}'_{ind}$ in Fig.~\ref{fig4b}, but local transfer term $-{\bf u}\cdot \nabla {\overline{\bf B}}$ is new.

%${\bf j}'_{2}$ in kinetic helicity and ${\bf j}_{2}$ in magnetic helicity are responsible for the quenching $\alpha$ effect. ${\bf u}_{pol}$ interacts with

\noindent ${\bf u}_{pol}$ interacts with $\int d\tau {\overline{\bf B}}\cdot \nabla {\bf u}$ to induce ${\bf j}'_{2}$, which yields the left handed helicity with ${\overline {\bf B}}_{pol}$. {\footnote {We add a subindex `$pol$' to ${\overline {\bf B}}$ to discriminate it from ${\overline {\bf B}}_{tor}$.}} ${\bf u}_{pol}$ also interacts with $\overline{\bf B}_{pol}$ (or ${\bf b}$) and $\int\,d\tau (-{\bf u}\cdot \nabla {\overline {\bf B}})$ yielding ${\bf j}'_{3}$ and ${\bf j}'_{4}$, respectively. The polarization of $\langle {\bf j}'_{3}\cdot \int d\tau \overline{\bf B}\cdot \nabla {\bf u}\rangle$ is opposite $(+)$ to that of $\langle {\bf j}'_2\cdot \overline{\bf B}_{pol}\rangle\,(-)$, but that of $\langle {\bf j}'_{4}\cdot \int d\tau \overline{\bf B}\cdot \nabla {\bf u}\rangle$ depends on the relative value of $-{\bf u}\cdot \nabla {\overline {\bf B}}$. When this locally transferred field is weak, ${\bf j'}_4$ is parallel to $\int d\tau ({\overline {\bf B}}\cdot \nabla {\bf u})$ producing the oppositely polarized ($+$) magnetic helicity with reference to $\langle {\bf j}'_2\cdot \overline{\bf B}_{pol}\rangle\,(-)$. However, as the strength of ${\overline {\bf B}}$ grows to surpass `$|{\bf b}|$', `$-{\bf u}\cdot \nabla {\overline {\bf B}}$' turns over so that the direction of ${\bf j}'_4$ is opposite to $\int {\overline {\bf B}}\cdot \nabla {\bf u}\,d\tau$ yielding the left handed $(-)$ magnetic helicity. This is the result of magnetic back reaction. However, this effect is negligible in HMFD where the helical magnetic field is continuously provided by the external source. The net magnetic helicity is `$+|\langle {\bf j}_2\cdot {\bf b}_{ind}\rangle|-|\langle {\bf j}_3\cdot {\bf b}_{pol}\rangle|-|\langle {\bf j}'_2\cdot {\overline {\bf B}}_{pol}\rangle|+|\langle {\bf j}'_3\cdot \int d\tau \overline{\bf B}\cdot \nabla {\bf u}\rangle|\pm|\langle {\bf j}'_{4}\cdot \int d\tau \overline{\bf B}\cdot \nabla {\bf u}\rangle|$'. The first four terms correspond to $\alpha$ effect, and the last term representing $\beta$ effect substantially amplifies the large scale magnetic field when $\alpha$ effect becomes negligible (Fig.~\ref{fig3a}, \ref{fig3b}).

\subsection{Analytical derivation of $\alpha$\&$\beta$}
In the helical dynamo, the basis of Eq.~(\ref{LS_magnetic_induction_equation_alpha_beta}) from Eq.~(\ref{LS_magnetic_induction_equation_raw}) is the replacement of small scale EMF $\langle \mathbf{u}\times \mathbf{b}\rangle$ with $\alpha \overline{\mathbf{B}}-\beta\nabla\times \overline{\mathbf{B}}$. This relation can be approximately derived using sort of a function iterative method with some appropriate closure theories such as  MFT\citep{Blackman}, EDQNM\cite{Pouquet}, DIA\citep{Yoshizawa}. All theories show qualitatively the same results; but, they have their own limitations, too. For example, for MFT, the variable $X$ is divided into the mean (large) scale quantity $\overline{\bf X}$ and small (turbulent) one $\bf x$. Then, they are taken average over the large scale $\langle \cdot \rangle$, and calculated with Reynolds rule and tensor identity.
\begin{eqnarray}
\alpha&=&\frac{1}{3}\int^t \big(\langle {\bf j}\cdot {\bf b}\rangle-\langle {\bf u}\cdot \nabla\times {\bf u}\rangle\big)\, d\tau,
\label{Mean_alpha}\\
\beta&=&\frac{1}{3}\int^t \langle u^2\rangle\, d\tau.
\label{Mean_beta}
\end{eqnarray}
During the analytic calculation, some turbulent variables with the triple correlation or higher order terms are derived. They are dropped with Reynolds rule or simply ignored with the assumption of being small. This may cause the increasing discrepancy between the real system and MFT as $Re_M$ grows. Also, the exact range of small scale regime and the eddy turnover time `$t$'  with integration are not yet known. If the whole turbulent scale regime is included, the large scale magnetic field with these $\alpha$\&$\beta$ exceeds the actual quantity. The small scale range for $\alpha$\&$\beta$ seems to be limited near to forcing scale \citep{Park2012a}. However, at present, there is no general way to calculate them except some simple dimensional analysis or experimental approach.\\

\noindent Another issue is the existence of large scale plasma motion $\overline{\bf U}$. If ${\bf U}\times {\bf B}$ is averaged over large scale and applied with Reynolds rule, two terms remain: $\xi\sim {\overline {\bf U}}\times {\overline {\bf B}}+\langle {\bf u}\times {\bf b}\rangle$. In principle, they should be replaced by $\alpha{\overline{\bf B}-\beta\nabla\times{\overline{\bf B}}}$. But, ${\overline {\bf U}}\times {\overline {\bf B}}$ is usually excluded with Galilean transformation. However, ${\overline {\bf U}}$ in simulation and observation does not disappear, rather its effect can grow with the increasing ${\overline{\bf B}}$. Eq.~(\ref{Mean_alpha}), (\ref{Mean_beta}) are over simplified results.\\

\noindent In DIA, those issues are included in formal Green's function $G$ with statistical second order relation.
\begin{eqnarray}
\langle X_i({k})X_j({-k})\rangle=(\delta_{ij}-\frac{k_ik_j}{k^2} )E_X(k)+\frac{i}{2}\frac{k_l}{k^2}\epsilon_{ijl}H_X(k)\label{Statistical_relation}\\
\bigg(\langle X^2\rangle=2\int E_X(k)d{k},\,\,\langle {\bf X}\cdot \nabla\times {\bf X}\rangle=\int H_X(k)d{k}\bigg)\nonumber
\end{eqnarray}
%($\langle X^2\rangle=2\int E_X(k)d{k}$, $\langle {\bf X}\cdot \nabla\times {\bf X}\rangle=\int H_X(k)d{k}$, and $|{\bf k}|=k$, $X=U\,or\, B$).
\noindent And, $\alpha$\&$\beta$ in DIA are
\begin{eqnarray}
\alpha&=&\frac{1}{3}\int d{\bf k}\int^t (G_M\langle {\bf j}\cdot {\bf b} \rangle-G_K\langle {\bf u}\cdot \nabla\times {\bf u}\rangle) d\tau, \label{DIA_alpha}\\
\beta&=&\frac{1}{3}\int^t  (G_K\langle u^2\rangle+G_M\langle b^2\rangle) d\tau.
\label{DIA_beta}
\end{eqnarray}
\noindent They are quite similar to those of MFT except $G_K$\&$G_M$ and turbulent magnetic energy $\langle b^2\rangle$ in $\beta$. $\alpha$ coefficient implies its quenching as $G_M\langle {\bf j}\cdot {\bf b} \rangle\rightarrow G_K\langle {\bf u}\cdot \nabla\times {\bf u}\rangle$. Also, the $\beta$ effect depends on the turbulent energy including $b^2$. Since DIA calculates kinetic approach and counter kinetic (magnetic) one separately, Eq.~(\ref{Statistical_relation}) yields $\langle u^2\rangle$ and $\langle b^2\rangle$ in $\beta$ (${\bf x}={\bf u},\,{\bf b}$).\\

\noindent $\alpha$\&$\beta$ calculated with EDQNM approximation show more or less similar physical properties such as quenching $\alpha$ and energy dependent positive $\beta$ \citep{Pouquet}.
\begin{eqnarray}
\alpha&=&\frac{2}{3}\int^{\infty} \Theta_{kpq}(t)\big(\langle {\bf j}\cdot {\bf b} \rangle- \langle {\bf u}\cdot \nabla\times {\bf u}\rangle\big)\,dq,
\label{EDQNM_alpha}\\
\beta&=&\frac{2}{3}\int^{\infty} \Theta_{kpq}(t)\langle u^2 \rangle\,dq,
\label{EDQNM_beta}
\end{eqnarray}
\noindent where triad relaxation time $\Theta_{kpq}=(1-exp(-\mu_{kpq}t))/\mu_{kpq}$ and eddy damping operator $\mu_{kpq}$ are used. {Note that $\Theta\sim \mu^{-1}_{kpq}$, $\mu_{kpq}=const.$  as $t\rightarrow \infty$ (see \cite{Pouquet} and references therein).} Formally, DIA or EDQNM is free from the nonlinear effects neglected in MFT. However, still there are unknown Green function, triad relaxation time, and eddy damping rate including eddy turnover time.\\

\noindent Furthermore, the small scale EMF $\langle \mathbf{u}\times \mathbf{b}\rangle$ used in two scale MFT and DIA is not well defined quantity. It is inferred from ${\bf X}-\overline{\bf X}$ which is supposed to be in the range of $k\ge 2$ in Fourier space. However, the range participating in the amplification of large scale field is very narrow. Our previous work to find $\alpha$\&$\beta$ with the conventional MFT shows that $\bf u$\&$\bf b$ (or $\alpha$\&$\beta$) exist only around the forcing scale. The whole turbulent range yields much larger growth of $\overline{\bf B}$ than actual value (Fig.~1, \cite{Park2012b}, Fig.~1b, \cite{Park2017}). Kolmogorov's inertia range seems to separate the range of $\bf u$\&$\bf b$ for $\alpha$\&$\beta$ from other dissipation scale. However, it is not clear whether the latter just dissipates or plays some other roles in dynamo. At least, they do not amplify $\overline{\bf B}$ directly. But, since the exact range cannot be found with theory, we may question if $\alpha$\&$\beta$ (or $\bf u$\&$\bf b$) are just conceptual quantities. Statistically, however, it makes sense to substitute $\alpha$\&$\beta$ and $\overline {\bf B}$ for EMF. And, the result is associated with the statistical correlation Eq.~(\ref{Statistical_relation}). If we apply \cite{Moffatt1978}'s assumption $EMF_i\sim\alpha_{ij}\overline{\mathbf{B}}_{j}+\beta_{ijk}\nabla_{k}\overline{\mathbf{B}}_j$\footnote{$\gamma$ is neglected for simplicity} to magnetic induction equation, we get
\begin{eqnarray}
\frac{\partial \overline{\mathbf{B}}}{\partial t}
\sim \nabla\times (\alpha \overline{\mathbf{B}}-(\beta+\eta) \nabla\times \overline{\mathbf{B}}).
\label{Large_B_for_statistics}
\end{eqnarray}
\noindent This shows that the growth rate of $\overline{\mathbf{B}}$ is represented by its curl effect. And, if the divergenceless magnetic field $\nabla\cdot \overline{\mathbf{B}}=0$ is added, this equation is led to Helmholtz theory. This means that magnetic induction equation with $\alpha$, $\beta$, and diffusion is statistically self consistent. The properly found $\alpha$ \& $\beta$ can describe the evolution of large scale magnetic field.\\% Since the equation represents the growth rate of a vector, not the second order moment in the stationary state, we can give the equation some flexibility with pseudo $\alpha$\& $\beta$.\\

\noindent This formal equation can be applied to the practical dynamo phenomenon such as Solar dynamo. If the equation is  divided into the poloidal component and toroidal one, two coupled equations from magnetic induction equation are derived \citep{Charbonneau}:
\begin{eqnarray}
&&\frac{\partial \overline{A}}{\partial t}=(\eta+\beta)\bigg(\nabla^2-\frac{1}{\varpi^2} \bigg)\overline{A}-\frac{{\bf u}_p}{\varpi}\cdot \nabla (\varpi\overline{A})+\alpha \overline{B}_{tor},\label{Solar_poloidal_magnetic_field}\\
&&\frac{\partial \overline{B}_{tor}}{\partial t}=(\eta+\beta)\bigg(\nabla^2-\frac{1}{\varpi^2} \bigg)\overline{B}_{tor}+\frac{1}{\varpi}\frac{\partial(\varpi\overline{B}_{tor})}{\partial r}\frac{\partial (\eta+\beta)}{\partial r}\label{Solar_toroidal_magnetic_field}\\
&&-\varpi{\bf u}_p\cdot \nabla \bigg(\frac{\overline{B}_{tor}}{\varpi}\bigg)-\overline{B}_{tor}\nabla\cdot{\bf u}_p+\varpi(\nabla \times (\overline{A}\hat{e}_\phi))\cdot\nabla{\bf \Omega}\nonumber\\
&&+\nabla\times(\alpha\nabla \times(\overline{A}\hat{e}_\phi)),\nonumber
\end{eqnarray}
\noindent where $\overline{\bf B}_{pol}=\nabla \times \overline{\bf A}$, $\varpi=r\,sin \theta$, and $\Omega$ is the angular velocity from convetive motion ${\overline{\bf U}={\bf r}\times {\bf \Omega}}$. This equation set reproduces the periodic solar magnetic field: amplification-annihilation-reverse. In Appendix, we show Solar dynamo simulation. Fig.~\ref{fig7a} in appendix includes the reproduction of \cite{Jouve}'s work. $\alpha$\&$\beta$ were chosen for the critical dynamo yielding 16.3 year period. Stefani et al.\cite{Stefani} added the effect of synchronized helicity oscillation from planets to $\alpha$ effect and solved the 1D equation to get $\sim 22$ year oscillation period. We solved it in 2D ($r,\,\theta$) simulation in spherical coordinates. Fig.~ \ref{fig7b} in appendix shows that the modified $\alpha$ reproduces the period of 21.7 without tuning any code variable. If additional physical effect exists, it can be applied to $\alpha$\&$\beta$ rather than EMF. It is also possible to infer $\alpha$\&$\beta$ from $E_M(t)$ and $H_M(t)$.\\

\begin{figure*}
\centering{
   \subfigure{
     \includegraphics[width=7.8 cm]{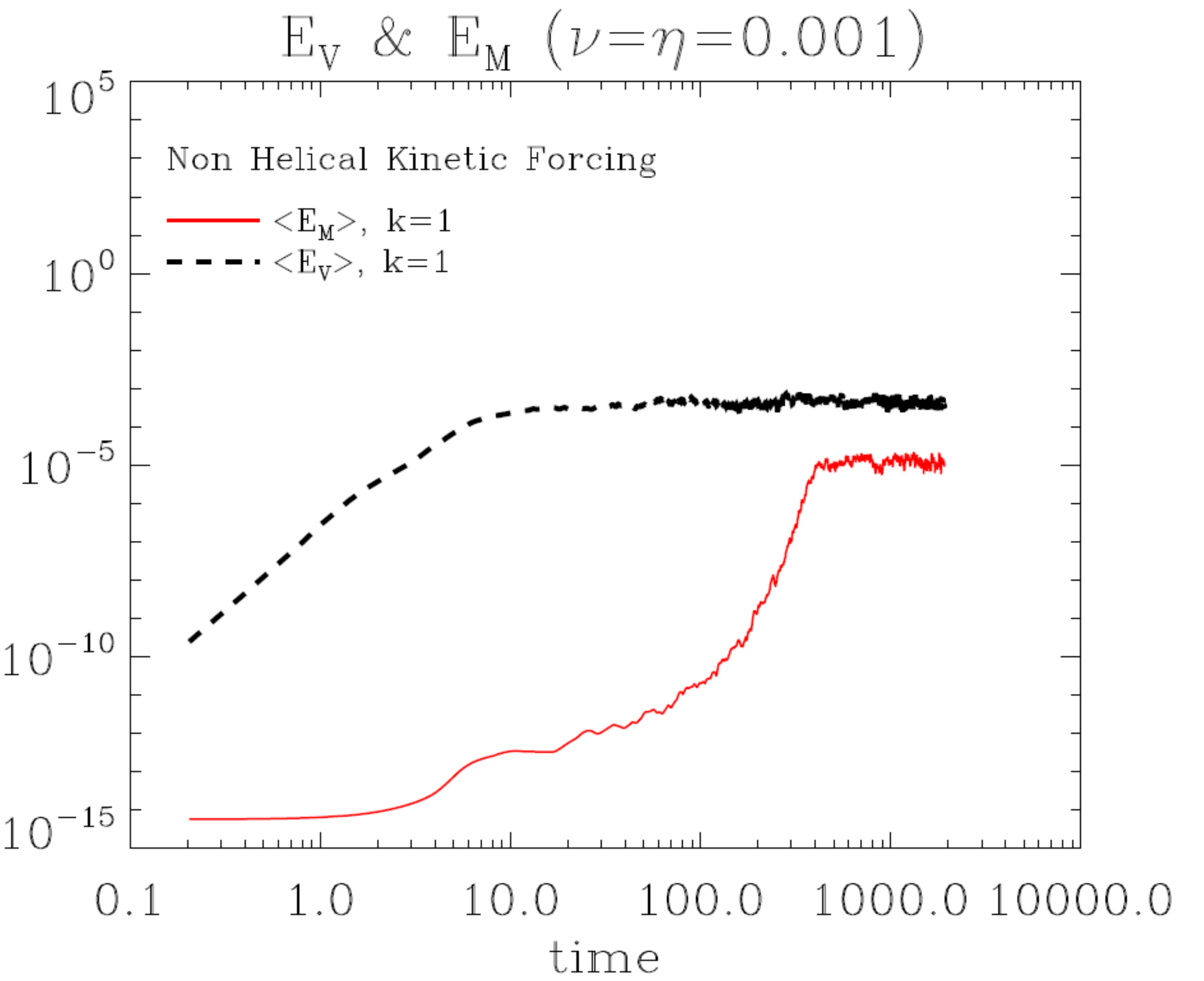}
     \label{fig6a}
}\hspace{-4 mm}
   \subfigure{
     \includegraphics[width=7.5 cm]{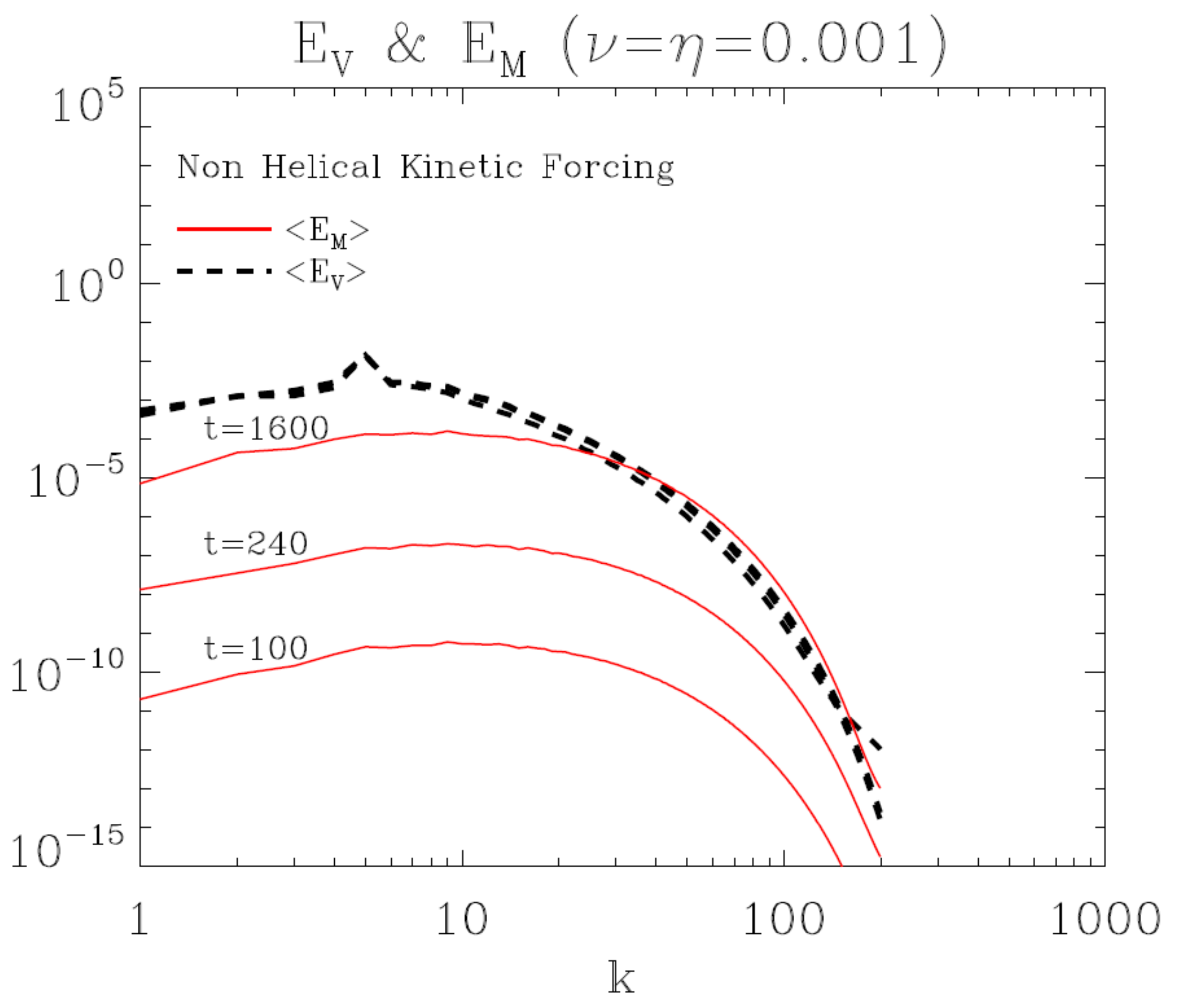}
     \label{fig6b}
   }
}
\caption{Nonhelical kinetic forcing small scale dynamo. Nonhelical velocity field was given to k=5. $\eta=\nu=0.001$, and magnetic Reynolds number $Re_M$ is $\sim 80$. Left panel shows the temporal evolution of $E_V$ and $E_M$. Rght panel shows their spectra at $t=100$, $240$, $1600$. The peak of $E_M$ is located between the forcing scale and dissipation scale.}
\end{figure*}

\subsection{Half Analytic and Half Numerical Method}
\noindent We can find $\alpha$ \& $\beta$ using the data of magnetic helicity and magnetic energy. We pointed out that Eq.~(\ref{Large_B_for_statistics}) is formally consistent with the statistical relation. Then, from Eq.~(\ref{LS_magnetic_induction_equation_alpha_beta}) we get
\begin{eqnarray}
\overline{\mathbf{B}}\cdot \frac{\partial \overline{\mathbf{B}}}{\partial t}
&=& \alpha \overline{\mathbf{B}}\cdot \nabla\times \overline{\mathbf{B}}+(\beta+\eta) \overline{\mathbf{B}}\cdot \nabla^2\overline{\mathbf{B}}\nonumber\\
&=& \alpha \overline{\mathbf{J}}\cdot \overline{\mathbf{B}}-(\beta+\eta)\, \overline{\mathbf{B}}\cdot \overline{\mathbf{B}}\,(k=1)\nonumber\\
\rightarrow \frac{\partial}{\partial t}\overline{E}_M&=& -2(\beta+\eta) {\overline E}_M+\alpha {\overline H}_M.
\label{Em1}
\end{eqnarray}
\noindent We can also derive the evolving magnetic helicity as follows.
\begin{eqnarray}
\rightarrow \frac{d}{dt}\overline{H}_M&=&4\alpha \overline{E}_M-2(\beta+\eta)\overline{H}_M.\label{Hm1}
\end{eqnarray}
\noindent These two equations are functions of actual data $\overline{E}_M$ and $\overline{H}_M$ resulting from all internal and external effects. One of a simple method to solve this coupled equation set is to diagonalize the matrix using a coefficient matrix $P$,which satisfies
$[{\overline H}_M,\, {\overline E}_M]\equiv P\,[X,\, Y]$. Then, we get\\
$\left[
\begin{array}{c}
\partial X / \partial t \\ %\frac{\partial X}{\partial t} \\
\partial Y / \partial t %\frac{\partial Y}{\partial t}
\end{array}
\right]
= P^{-1}
\left[
\begin{array}{cc}
-2(\beta+\eta) & 4\alpha\\
     \alpha         &-2(\beta+\eta)
\end{array}
\right]
P
\left[
\begin{array}{c}
X \\
Y
\end{array}
\right]
=
\left[
\begin{array}{cc}
\lambda_1 & 0\\
0 &\lambda_2
\end{array}
\right]
\left[
\begin{array}{c}
X \\
Y
\end{array}
\right]. $\\

\noindent $[{\overline H}_M,\, {\overline E}_M]$ can be found from $P^{-1}[X,\, Y]$, where the column vector of $P$ forms the basis of  eigenvectors. The result is,
\begin{eqnarray}
2\overline{H}_M(t_n)&=&(2\overline{E}_M(t_{n-1})+\overline{H}_M(t_{n-1}))e^{2\int^{t_n}_0(\alpha-\beta-\eta)d\tau}\nonumber\\
&&-(2\overline{E}_M(t_{n-1})-\overline{H}_M(t_{n-1}))e^{2\int^{t_n}_0(-\alpha-\beta-\eta)d\tau},\label{HmSolutionwithAlphaBeta1}\\
4\overline{E}_M(t_n)&=&(2\overline{E}_M(t_{n-1})+\overline{H}_M(t_{n-1}))e^{2\int^{t_n}_0(\alpha-\beta-\eta)d\tau}\nonumber\\
&&+(2\overline{E}_M(t_{n-1})-\overline{H}_M(t_{n-1}))e^{2\int^{t_n}_0(-\alpha-\beta-\eta)d\tau}.\label{EmSolutionwithAlphaBeta2}
\end{eqnarray}
$\overline{H}_M$ is always smaller than $2\overline{E}_M$, which satisfies realizability condition. But, $\overline{H}_M\rightarrow 2\overline{E}_M$ as the system is getting saturated. In case of right handed HMFD, clearly $\alpha>0$ so that the first term in Eq.~(\ref{HmSolutionwithAlphaBeta1}), (\ref{EmSolutionwithAlphaBeta2}) are dominant. This means $\overline{H}_M(t_n)$ as well as $\overline{E}_M(t_n)$ is positive. But in case of left handed HMFD, the second term is dominant. This indicates that  $\overline{H}_M(t_n)$ is negative, but $\overline{E}_M(t_n)$ is positive. On the contrary, in case of positively forced HKFD, $\alpha$ is negative so that the second term in each equation is dominant leading to negative $\overline{H}_M$. Still, $\overline{E}_M$ is not influenced by  the chirality of forcing. These inferences are well consistent with the simulation result of HKFD or HMFD.\\

\noindent $\alpha$ \& $\beta$ from above results are \citep{Park2020}
\begin{eqnarray}
\alpha(t)&=&\frac{1}{4}\frac{d}{dt}log_e \bigg|\frac{ 2\overline{E}_M(t)+\overline{H}_M(t)}{2\overline{E}_M(t)-\overline{H}_M(t)}\bigg|,\label{alphaSolution3}\\
\beta(t)&=&-\frac{1}{4}\frac{d}{dt}log_e\big| \big(2\overline{E}_M(t)-\overline{H}_M(t) \big)\big( 2\overline{E}_M(t)+\overline{H}_M(t)\big)\big|-\eta.\nonumber\\
\label{betaSolution3}
\end{eqnarray}
\noindent To get the $\alpha$\&$\beta$, we need the simulation or observation data of $\overline{E}_M(t)$ and $\overline{H}_M(t)$ in each time `$t_n$'. For example, $d\overline{E}_M/dt$ is approximately $ \sim (\overline{E}_M(t_n)-\overline{E}_M(t_{n-1}))/(t_n-t_{n-1})$. We compared $\nabla \times \langle {\bf u}\times {\bf b}\rangle$ with $\nabla \times (\alpha {\overline{\bf B}}-\beta\nabla \times {\overline{\bf B}})$ in Fig.~\ref{fig3a1}, \ref{fig3b1}. In the early time regime, they are quite close to each other. But, oscillation increases as the field becomes saturated ($2\overline{E}_M\sim\overline{H}_M$). Also, note that we used $k=1$ for the large scale field in the analytic and numerical calculation. And, for the anisotropic system, we need data for $\overline{E}_{\bot,\,M}(t)$ \& $\overline{H}_{\bot,\,M}(t)$, $\overline{E}_{||,\,M}(t)$ \& $\overline{H}_{||,\,M}(t)$, and  the anisotropic solution of Eq.~(\ref{alphaSolution3}), (\ref{betaSolution3}). When these theoretical results are applied to the real data, some appropriate filtering, simplifying, or normalizing data may be required according the quality of data.
%In our case, only the simple filtering process was used.

\subsection{Derivation of $\beta$}
\noindent Now, we check the possibility of negative $\beta$ using analytic method.
%\begin{eqnarray}
%&&{\bf u}\times (-{\bf u}\cdot \nabla \overline{\bf B})\rightarrow-\epsilon_{ijk}\langle u_j (r,\,t)u_m(r+l,\,\tau)\rangle\frac{\partial \overline{B}_k}{\partial \overline{r}_m}\label{beta_derivation_helical_first}\\
%&\sim&-\epsilon_{ijk}\langle u_j(t)u_m(\tau)\rangle\frac{\partial \overline{B}_k}{\partial \overline{r}_m}-\langle u_j(t)\,l_n\partial_n %u_m(\tau)\rangle\epsilon_{ijk}\frac{\partial \overline{B}_k}{\partial \overline{r}_m}\\
%&\sim&\underbrace{-\frac{1}{3}\langle u^2\rangle\epsilon_{ijk}\frac{\partial \overline{B}_k}{\partial \overline{r}_m}\delta_{jm}}_1\,\underbrace{-\big \langle \epsilon_{jnm}\frac{l}{6}|H_V|\big \rangle\epsilon_{ijk}\frac{\partial \overline{B}_k}{\partial \overline{r}_m}\delta_{nk}\delta_{mi}}_2,\label{beta_derivation_helical1}
%\end{eqnarray}
\begin{eqnarray}
&&\langle {\bf u}\times (-{\bf u}\cdot \nabla \overline{\bf B})\rangle \rightarrow \big\langle -\epsilon_{ijk} u_j (r,\,t)u_m(r+l,\,\tau)\frac{\partial \overline{B}_k}{\partial \overline{r}_m}\big\rangle \label{beta_derivation_helical_first}\\
&\sim&-\epsilon_{ijk}\langle u_j(t)u_m(\tau)\rangle\frac{\partial \overline{B}_k}{\partial \overline{r}_m}-\langle u_j(t)\,l_n\partial_n u_m(\tau)\rangle\epsilon_{ijk}\frac{\partial \overline{B}_k}{\partial \overline{r}_m}\\
&\sim&\underbrace{-\frac{1}{3}\langle u^2\rangle\epsilon_{ijk}\frac{\partial \overline{B}_k}{\partial \overline{r}_m}\delta_{jm}}_1\,\underbrace{-\big \langle \frac{l}{6}|H_V|\big \rangle\epsilon_{ijk}\frac{\partial
%\overline{r}_m}\delta_{jm}}_1\,\underbrace{-\big \langle \epsilon_{jnm}\frac{l}{6}|H_V|\big \rangle\epsilon_{ijk}\frac{\partial
\overline{B}_k}{\partial \overline{r}_m}\delta_{nk}\delta_{mi}}_2,\label{beta_derivation_helical1}
\end{eqnarray}
\noindent {In Eq.~(\ref{beta_derivation_helical_first}) we used the approximation of $(u(r+l,\,\tau)-u(r,\,\tau))/l \equiv \Delta u / \Delta r \sim \partial u(r',\,\tau)/\partial r$} with the spatially small increment `$l$'. We assumed that the sufficiently many eddies make the plasma system statistically continuous. Then, there should be at least one point `$r'$' between `$r+l$' and `$r$', whose tangential line is parallel to the slope between them. Moreover, it is not difficult to find $\partial u/\partial r > u(r+l)$, especially when $u(r+l)u(r)<0$.\\

\noindent For the nonhelical field, the result will be $(\langle u^2/3\rangle+l\cdot\nabla \langle u^2/2\rangle)(-\nabla \times \overline{\bf{B}})$, which is the conventional positive $\beta$. {For the helical fields of ${u}_j$, the first term can be still calculated as the energy density `$m\rightarrow j$'.} However, in the second term, $u_j$, $u_m$, and $\overline{B}_k$  are mutually perpendicular. Without loss of generality, `${u}_j$' can be considered as `${u}_{pol}$'. Then, `${u}_m$' should be `${u}_{tor}$' (`$m$'$\rightarrow$`$i$'), and `$n$' should be `$k$'.\footnote{Kronecker delta is used to indicate the index for a specific direction. It is not from Levi-Civita relation. Also, note that $\partial \overline{B}_k/\partial \overline{r}_m$ is taken out of the bracket which is for the average over large scale.} Since the turbulent velocity part is saturated earlier than the large scale eddy, {it can be calculated separately: $l/6\langle {\bf u}\cdot \nabla\times {\bf u}\rangle(\equiv l/6H_V)$.} {As mentioned, `$l_n$' or `$l(=|l_n|)$' is the increment of `$r$'. But it can be also considered as correlation length. Since the helical velocity field ${\bf u}_{tor}$ can be converted into ${\bf u}_{pol}$ through curl operator, `$l_n$' indicates lateral correlation length.} For the left handed helical structure, we first place a virtual mirror on the right side of ${\bf u}_{tor,\,2}$. Then, the toroidal velocity eddies will be reflected on the other side, and the correlation length `$l_n$' is toward `$-\hat{l}$'. The reflection makes $\nabla_n {u}_m$ negative, but ${\bf u}_{pol}$ does not change. That is, `$l_n$' should be sort of a pseudo-scalar (refer to the difference between $\langle u^2\rangle$ and $\langle {\bf u}\cdot {\bf \omega}\rangle$). Considering the statistical meaning of each component, we can make a more general form as follows:
\begin{eqnarray}
%&\Rightarrow&-\int^t \frac{1}{3}  \langle u^2\rangle d\tau\,(\nabla \times \overline{\bf B})+ \int^t  \frac{l}{6} |H_V|\, d\tau\,(\nabla \times \overline{\bf B}).
%&\Rightarrow&-\bigg(\int^t \frac{1}{3}  \langle u^2\rangle d\tau - \int^t  \frac{l}{6} |H_V|\, d\tau\bigg) \nabla \times \overline{\bf B}_j.
\sim-\frac{1}{3}\langle u^2\rangle \nabla\times \overline{\bf B}+\frac{l}{6} |H_V|\nabla \times \overline{\bf B}\rightarrow -\beta\nabla \times \overline{\bf B}.
%&\Rightarrow&-\big(\beta_1(t)+\beta_2(t)\big) \nabla \times \overline{\bf B}.
\label{beta_derivation_helical2}
\end{eqnarray}
The total diffusion effect becomes $(\beta+\eta)\nabla^2{\overline {\bf B}}$, whose Fourier transformed expression is $-(\beta+\eta)k^2{\overline {\bf B}}$ regardless of chirality.\\

%If we use a simple dimensional analysis leading to $l\partial /\partial r\sim 1$, we get $u(r+l)\sim u(r) + u(r)+1/2u(r)...$. But this is not a generally accepted case.

\noindent With these analytical analysis, `$l$' is not yet clearly defined.  If we consider the correlation length between two eddies (helical field), we can derive with the parallel correlation $g(r)$
\begin{eqnarray}
\langle u(r)u(r+l)\rangle\equiv \langle u^2(r)\rangle g(r)\sim \frac{1}{3}\langle u^2\rangle-\frac{l}{6}\langle {\bf u}\cdot \nabla\times {\bf u}\rangle\sim \frac{\big(2-lk \big)}{6}\langle u^2\rangle.
\label{new_beta_derivation_helical3}
\end{eqnarray}
\noindent The condition for $2-lk<0$ is actually the correlation position that makes $g(r)<0$, which is the typical property of parallel correlation length\citep{Davidson}. We can get the condition $l>2/k$. Since $k$ for small scale is larger than 2, the condition of negative $\beta$ is not the hard one. At present, we leave $lH_V/6$ to be an independent quantity. But it should be noted that this term cannot be considered as $\alpha$ effect because $\nabla \times (-\langle u^2\rangle/3+l/|6H_V|) \nabla \times \overline{\bf B}\rightarrow (\langle u^2\rangle/3-l/6|H_V|) \nabla^2 \overline{\bf B}$. This is a typical diffusion term.\\

%On the other hand, if we just apply the dimension of $l$ to the calculation, we just get  $l\partial/\partial r\sim 1$ leading to $u(r+l)-u(r)\sim u(r)\rightarrow u(r+l)\sim 2u(r)$. But it is not a general case. It should be careful for applying the dimensional analysis to finding the quantitative condition.\\

\noindent Fig.~\ref{fig8} in appendix includes how $1/3\langle u^2\rangle$ and $1/6\langle {\bf u}\cdot \nabla\times {\bf u}\rangle$ evolve. It is easy to check $\langle {\bf u}\cdot \nabla\times {\bf u}\rangle\sim k\langle u^2\rangle$ with $k=$2, 3, 4, ... is larger than $\langle u^2\rangle$ both in Fourier and real space. This negative magnetic diffusivity plays and opposite role in plasma motion (see Eq.~(\ref{negative_beta_compressing_plasma_motion1}), (\ref{negative_beta_compressing_plasma_motion2})).\\

\noindent On the other hand, Kraichnan derived the negative magnetic diffusion effect in Lagrangian formation with the assumption of strong helical field(\cite{Kraichnan}, see also \cite{Brandenburg_Sokoloff}):
\begin{eqnarray}
\frac{\partial \overline{\mathbf{B}}}{\partial t}=\beta_0\nabla^2\overline{\mathbf{B}}+\tau_2\nabla\times \langle\alpha\nabla\times \alpha\rangle\,\overline{\mathbf{B}}\rightarrow(\beta_0-\tau_2A)\nabla^2\overline{\mathbf{B}}.
\label{Kraichnan_alpha_beta}
\end{eqnarray}
(Here, $\beta_0=\int^{\tau_1}u^2_0 dt\sim \tau_1u^2_0,\, \alpha(\mathbf{x},\, t)=(-)1/3\langle \mathbf{u}\cdot\omega \rangle\tau_1$, $\langle \alpha(\mathbf{x},\,t)\alpha(\mathbf{x}',\,t')\rangle$=$A(x-x')D_2(t-t'),\, \tau_2=\int^{\infty}D_2(t)\, dt$.)\\
$\beta_0$ is the conventional positive magnetic diffusion effect. But, $-\tau_2A$, which is from the correlation of $\langle \alpha \alpha\rangle$, actually plays the role of the negative magnetic diffusion $\beta$ effect. The detailed derivation of Eq.~(\ref{Kraichnan_alpha_beta}) is not the same as Eq.~(\ref{beta_derivation_helical_first})-(\ref{beta_derivation_helical1}). However, they both are from the turbulent velocity $\bf u$ and shows its helical feature produces the negative magnetic diffusion. Besides, there are some theoretical and experimental works associated with negative magnetic diffusivity (\citep{Kraichnan, Lanotte}, references therein). They are based on $\alpha-\alpha$ correlation in the strong helical system and do not explain the coupling of ${\overline H}_M$ and ${\overline E}_M$. These approaches are different from ours. However, it should be noted that the $\alpha-\alpha$ correlation from $\alpha$ effect eventually plays a role of negative magnetic diffusivity as the equation indicates.\\

\noindent {\bf \noindent So far, we have argued that the main reason of negative magnetic diffusivity is helical component in `$u$'(see Eq.~(\ref{beta_derivation_helical2}), (\ref{Kraichnan_alpha_beta})). We can refer to the numerically supporting result in Fig.~\ref{fig9} in appendix \cite{Park2020}. When the helical kinetic forcing is turned off at $t\sim 200$ and nonhelical forcing is on, negative $\beta$ becomes positive. This negative magnetic diffusivity suppresses the growth of large scale magnetic field. It should be noted that $\beta$ is the function of turbulent kinetic energy and kinetic helicity, not the forcing method itself.}\\

\noindent {\bf The negative magnetic diffusivity is observed in liquid sodium experiment\cite{Simon et al}. Simon et al. found that the small scale turbulent fluctuations ($\sim u$) contribute to the negative magnetic diffusivity in the interior region. $\alpha$ effect practically disappears for the lower $Re_M$ value. Instead, $\beta$ effect increases strongly, which can promote the dynamo action.}

%$\alpha$ in Eq.~(\ref{beta_derivation_helical_first}) is from $\langle {\bf u}\cdot {\bf \omega}\rangle$ without current density $\langle {\bf j}\cdot {\bf b}\rangle$.

% ($\langle b\rangle=\langle bbb\rangle=0$, $\langle bb\rangle\neq 0$...)
\subsection{Plasma quenching}
As Fig.~1(a), 1(b) show, we briefly discuss how the helical magnetic field constrains plasma with the negative $\beta$ effect \citep{Park2020}. As Eq.~(\ref{momentum_equation_original}), (\ref{magnetic_induction_equation_original}) imply, plasma and magnetic field are coupled through Lorentz force and EMF. If we take the scalar product of $\bf B$ or $\bf U$ on the curl of EMF or momentum equation respectively, we get ${\bf B}\cdot\nabla\times ({\bf U}\times {\bf B})$ or ${\bf U}\cdot({\bf J}\times {\bf B})$. And they are practically the same except the opposite sign. To make it clear,  the field scales can be divided into $\overline{\bf U}$, $\overline{\bf B}$ and turbulent $\bf u$, $\bf b$. Taking the average and applying Reynolds rule, we get
\begin{eqnarray}
{{\bf U}}\cdot {\bf J}\times {\bf B}&=&-{\overline{\bf B}}\cdot\nabla\times (\overline{\bf U}\times \overline{\bf B})-\overline{\bf B}\cdot\nabla\times \langle{\bf u}\times {\bf b}\rangle\nonumber\\
&&-\langle{\bf b}\cdot\nabla\times (\overline{\bf U}\times {\bf b})\rangle-\langle{\bf b}\cdot\nabla\times {\bf u}\times \overline{\bf B}\rangle
\label{negative_beta_compressing_plasma_motion1}
\end{eqnarray}
\noindent Considering Fig.~2, we see the first term is negligible. The third term is not so significant because of the high  helicity ratio in the small scale regime. And the fourth term can be dropped replacing ${\bf j}$ with $\rho\, {\bf u}$. The second term  can be rewritten like
\begin{eqnarray}
-\overline{\bf B}\cdot\nabla\times \langle{\bf u}\times {\bf b}\rangle&=&-\overline{\bf B}\cdot\nabla\times(\alpha \overline{\bf B}-\beta \nabla \times \overline{\bf B})\nonumber\\
&=&-\alpha\overline{\bf B}\cdot\nabla\times\overline{\bf B}-\beta \overline{\bf B}\cdot \nabla^2 \overline{\bf B}.
\label{negative_beta_compressing_plasma_motion2}
\end{eqnarray}
\noindent The first term becomes negligible, but the second term is $-\beta \overline{\bf B}\cdot \nabla^2 \overline{\bf B}\rightarrow \beta k^2\overline{B}^2$. Then, the negative $\beta$ suppresses the plasma motion $\overline{\bf U}$ while it amplifies $\overline{\bf B}$ (see Fig.~1(a), (b)).\\

\noindent In comparison with helical large scale dynamo in Fig.~1, nonhelical small scale dynamo in Fig.~7 shows the supplementary role of $\beta$ in the large scale plasma motion $\overline {\bf U}$. Although small scale kinetic energy ($\sim \langle u^2\rangle$) is much larger than that of HMFD, there is no significant decrease in $E_V(\sim {\overline U^2}$). This indicates that the positive $\beta$ provides magnetic energy to the large scale plasma motion.\\

\section{Summary}
In this paper, we have pointed out the possibility of helical magnetic forcing dynamo (HMFD). HMFD has several features distinguished from helical kinetic forcing dynamo (HKFD). Externally given $E_M$ is converted into $E_V$ through Lorentz force, which activates the plasma motion  and EMF. This nontrivial EMF transports $E_M$ into the large and small scale region. $E_V$ in HMFD is subsidiary to the migration of $E_M$ so that magnetic Reynolds number $Re_M(=UL/\eta)$ is negligibly small. Large scale magnetic energy ${\overline E}_M$ is amplified and saturated more efficiently than that of HKFD. Compared to our previous HKFD experiment, only 20\% of provided magnetic energy produced the higher magnetic energy level. This may be able to explain the big gap between the cosmological seed magnetic field ($\sim 10^{-19} G$) and the galactic magnetic field ($\sim 10^{-6} G$).\\

%{\bf The correlation length of magnetic field is constrained by the scale of small horizon during inflation. It could have been expanded by some MHD process. However, without the inverse cascade of helical magnetic energy, the strength of current large scale magnetic field is hard to be explained.}\\

\noindent {The nonlinear interaction of helical field and plasma can be explained with $\alpha$ and $\beta$ effect which linearlize the nonlinear dynamo process. Since the exact definitions of $\alpha$\&$\beta$ are not yet known, we calculated them using Eq.~(\ref{alphaSolution3}), (\ref{betaSolution3}). Compared to the conventional theory, $\alpha$ {\bf becomes negligible} much earlier than the saturation of ${\overline {\bf B}}$. In contrast, $\beta$ keeps negative and gets saturated along with ${\overline E}_M$. Clearly, the effect of $\alpha$ as a generator of ${\overline E}_M$ is not much. Rather, the negative $\beta$ effect plays the substantial role of amplifying ${\overline E}_M$ with Laplacian $\nabla^2\rightarrow -k^2$(k=1).} And we have discussed the wavenumber for the normalized large scale field should be constantly 1. In Fig.~ 2, 4, we used k=1 leading to the consistent result.\\

\noindent {The main dynamo processes in the system are as follows: $\int \partial {\bf u}/\partial t\,d\tau \times {\bf b} \sim \int {\overline{\bf B}}\cdot \nabla {\bf b}\,d\tau\times {\bf b}\sim \int ({\bf b}\cdot \nabla \times {\bf b})\,d\tau\,\,{\overline{\bf B}}$ (positive magnetic helicity), ${\bf u}\times \int \partial {\bf b}/\partial t\,d\tau\sim {\bf u}\times \int {\overline{\bf B}}\cdot \nabla {\bf u}\,d\tau\sim -\int ({\bf u}\cdot \nabla \times {\bf u})\,d\tau \,{\overline{\bf B}}$ (negative magnetic helicity), ${\bf u}\times \int -{\bf u}\cdot \nabla {\overline{\bf B}}d\tau \times {\bf b} \sim \beta\nabla^2{\overline{\bf B}}$ (positive magnetic helicity). The first two interactions correspond to $\alpha$ quenching, and the last one is associated to the negative $\beta$ effect. These $\alpha$ and negative $\beta$ effect are commonly originated from the helical component in velocity field. In addition to the helical effect in conventional $\alpha$ effect, we newly showed that the role of helical velocity field in the advection term $-{\bf u}\cdot \nabla \overline{\bf B}$ leads to the negative $\beta$ effect. And its prerequisite is turbulent kinetic helicity is larger than kinetic energy. We prepared for a supporting plot in Fig.~\ref{fig8} in appendix.\\

% On the other hand, Moffatt \& Kraichnan's $\langle \alpha(t) \alpha(t-\tau)\rangle$ correlation plays the same role as negative magnetic diffusion effect. On the contrary, the nonhelical component in `$\bf u$' results in the conventional positive $\beta$ effect.

% Finally, Test Field Method also shows the possibility of negative $\beta$ effect (Fig.~ 7 in \cite{Warnecke}).

\noindent We showed the evolution of $\alpha$, $\beta$ in Fig.~3 and verified their consistency in Fig.~4. The importance of this approach is the separation of $\alpha$\&$\beta$ from EMF without ambiguity. Numerically and analytically verified $\alpha$\&$\beta$ give us some clues to solve the nonlinear effects neglected in their analytic derivations as well as their experimental application. As long as EMF is represented as $\alpha \overline{\bf B}-\beta\nabla \times \overline{\bf B}$, negative $\beta$ replacing the quenching $\alpha$ effect is a necessary condition for the amplification of large scale field. Moreover, the negative $\beta$ effect suppresses the large scale plasma motion in the helical system.\\

\noindent {\bf The physical feature and effect of helical magnetic field in plasma are useful to investigate the origin of PMF as well as the current astrophysical phenomena. Biermann battery effect shows how the seed magnetic field in the early universe could be generated. And, neutrino-lepton interaction is a promising candidate of magnetic helicity in the Universe. The inverse cascade of magnetic helicity gives us a clue to the expansion of PMF scale constrained by the small horizon during inflation and amplification of its strength. And macroscopically, magnetic helicity explains how the evolution of magnetic field in plasma is constrained, which leads to the evolution of astro-plasma system eventually. All these events are closely related to HMFD.\\}

\noindent In this paper we considered only the case of $Pr_M=1$ with fully helical field in HMFD. However, we need to study more general system with $Pr_M\neq 1$ and arbitrary helicity ratio. Especially, the generation of poloidal magnetic field from the toroidal field in the solar convection zone with such a low $Pr_M\, (\sim 10^{-2})$ challenges the current dynamo theory.}\\

%%%%%%%%%%%%%%%%%%%%%%%%%%%%%%%%%%%%%%%%%%%%%%%%%

\noindent{\bf Acknowledgements}\\
The authors appreciate support from National Research Foundation of Korea:
NRF-2021R1I1A1A01057517, NRF-2020R1A2C3006177, and NRF-2021R1A6A1A03043957. Also, K. Park appreciates Dr. Frank Stefani and Dr. Matthias Rheinhardt for their useful advice on the Solar dynamo simulation.\\

%\bibliography{bibfile} % if your bibtex file is called example.

\begin{thebibliography}{99}

%\bibitem{}{a}
%Author, \emph{Title}, \emph{J. Abbrev.} {\bf vol} (year) pg.

% Please avoid comments such as "For a review'', "For some examples",
% "and references therein" or move them in the text. In general,
% please leave only references in the bibliography and move all
% accessory text in footnotes.


\bibitem{Balbus}
Balbus S. A., Hawley J. F., 1991, ApJ, 376, 214
\bibitem{Machida}
Machida M. N., Matsumoto T., Tomisaka K., Hanawa T., 2005, MNRAS, 362, 369
\bibitem{Boyd}
Boyd T. J. M., Sanderson J. J., 2003, The Physics of Plasmas. Cambridge University Press
\bibitem{Martin}
Martin J., Yokoyama J., 2008, J. Cosmology Astropart. Phys., 2008, 025
\bibitem{Subramanian}
Subramanian K., 2016, Reports on Progress in Physics, 79, 076901
\bibitem{Yamazaki}
Yamazaki D. G., Kajino T., Mathews G. J., Ichiki K., 2012, Phys. Rep., 517, 141
\bibitem{Cheng}
Cheng B., Olinto A. V., 1994, Phys. Rev. D, 50, 2421
\bibitem{Tevzadze}
Tevzadze A. G., Kisslinger L., Brand enburg A., Kahniashvili T., 2012, ApJ, 759, 54
\bibitem{Harrison}
Harrison E. R., 1970, MNRAS, 147, 279
\bibitem{Biermann}
Biermann L., 1950, Zeitschrift Naturforschung Teil A, 5, 65
\bibitem{Semikoz_Sokoloff}
Semikoz V. B., Sokoloff D., 2005, A\&A, 433, L53
\bibitem{Yamazaki et al}
Dai G. Yamazaki and Motohiko Kusakabe, 2012, Phys. Rev. D 86, 123006
\bibitem{Luo et al}
Luo Yudong, Kajino Toshitaka, Kusakabe Motohiko, and Mathews Grant J., APJ, 2019, 872, 172
\bibitem{Moffatt1978}
Moffatt H. K., 1978, Magnetic field generation in electrically conducting fluids, Cambridge, England, Cambridge University Press, 1978. 353p
\bibitem{Pouquet}
Pouquet A., Frisch U., Leorat J., 1976, Journal of Fluid Mechanics, 77, 321
\bibitem{Yoshizawa}
Yoshizawa A., 2011, Hydrodynamic and Magnetohydrodynamic Turbulent Flows, 1998, Springer
\bibitem{Park2020}
Park K., 2020, ApJ, 898, 112
\bibitem{Brandenburg}
Brandenburg A., 2001, ApJ, 550, 824
\bibitem{Jouve}
Jouve L., et al., 2008, A\&A, 483, 949
\bibitem{Kraichnan}
Kraichnan R. H., 1976, Journal of Fluid Mechanics, 75, 657
\bibitem{Blackman}
Blackman E. G., Field G. B., 2002, Physical Review Letters, 89, 265007
\bibitem{Park2012a}
Park K., Blackman E. G., 2012a, MNRAS, 419, 913
\bibitem{Park2012b}
Park K., Blackman E. G., 2012b, MNRAS, 423, 2120
\bibitem{Park2017}
Park K., 2017, MNRAS, 472, 1628
\bibitem{Charbonneau}
Charbonneau P., 2014, ARA\&A, 52, 251
\bibitem{Stefani}
Stefani F., Giesecke A., Weber N., Weier T., 2016, Sol. Phys., 291, 2197
\bibitem{Davidson}
Davidson P. A., 2004, Turbulence : an introduction for scientists and engineers, 2004, OUP Oxford
\bibitem{Brandenburg_Sokoloff}
Brandenburg, A., Sokoloff, D., 2002, Geophys. Astrophys. Fluid Dyn., 96, 319
\bibitem{Lanotte}
Lanotte A., Noullez A., Vergassola M., Wirth A., 1999, Geophysical and Astrophysical Fluid Dynamics, 91, 131
\bibitem{Simon et al}
Simon Cabanes, Nathanaël Schaeffer, Henri-Claude Nataf, 2014, Phys. Rev. Letter, 113, 184501
\bibitem{Semikoz}
Semikoz V. B., 2004, arXiv e-prints, pp hep–ph/0403096
%\bibitem{Warnecke}
%Warnecke J., Rheinhardt M., Tuomisto S., Käpylä P. J., Käpylä M. J., Brandenburg A., 2018, A\&A, 609, A51


% Also, please have only one work for each \bibitem{}.
%\bibitem{Balbus}
%Balbus S. A., Hawley J. F., 1991, ApJ, 376, 214
%\bibitem{Biermann}
%Biermann L., 1950, Zeitschrift Naturforschung Teil A, 5, 65
%\bibitem{Blackman}
%Blackman E. G., Field G. B., 2002, Physical Review Letters, 89, 265007
%\bibitem{Boyd}
%Boyd T. J. M., Sanderson J. J., 2003, The Physics of Plasmas. Cambridge University Press
%\bibitem{Brandenburg}
%Brandenburg A., 2001, ApJ, 550, 824
%\bibitem{Brandenburg_Sokoloff}
%Brandenburg, A., Sokoloff, D., 2002, Geophys. Astrophys. Fluid Dyn., 96, 319
%\bibitem{Brandenburg_Subramanian}
%Brandenburg A., Subramanian K., 2005, Phys. Reports, 417, 1
%\bibitem{Charbonneau}
%Charbonneau P., 2014, ARA\&A, 52, 251
%\bibitem{Cheng}
%Cheng B., Olinto A. V., 1994, Phys. Rev. D, 50, 2421
%\bibitem{Davidson}
%Davidson P. A., 2004, Turbulence : an introduction for scientists and engineers
%%\bibitem{Devlen}
%%Devlen E., Brandenburg A., Mitra D., 2013, MNRAS, 432, 1651
%\bibitem{Harrison}
%Harrison E. R., 1970, MNRAS, 147, 279
%\bibitem{Jouve}
%Jouve L., et al., 2008, A\&A, 483, 949
%%\bibitem{KapylP}
%%Käpylä P. J., Korpi M. J., Brandenburg A., 2009, A\&A, 500, 633
%%\bibitem{KapylM}
%%Käpylä M. J., Vizoso J. Á., Rheinhardt M., Brandenburg A., Singh N. K., 2020, ApJ, 905, 179
%\bibitem{Kraichnan}
%Kraichnan R. H., 1976, Journal of Fluid Mechanics, 75, 657
%\bibitem{Krause}
%Krause F., Rädler K., 1980, Mean-field magnetohydrodynamics and dynamo theory. Oxford,
%Pergamon Press, Ltd., 1980. 271 p.
%\bibitem{Lanotte}
%Lanotte A., Noullez A., Vergassola M., Wirth A., 1999, Geophysical and Astrophysical Fluid Dynamics, 91, 131
%\bibitem{Luo et al}
%Luo Yudong, Kajino Toshitaka, Kusakabe Motohiko, and Mathews Grant J., APJ, 2019, 872, 172
%\bibitem{Machida}
%Machida M. N., Matsumoto T., Tomisaka K., Hanawa T., 2005, MNRAS, 362, 369
%\bibitem{Martin}
%Martin J., Yokoyama J., 2008, J. Cosmology Astropart. Phys., 2008, 025
%\bibitem{Moffatt1974}
%Moffatt H. K., 1974, Journal of Fluid Mechanics, 65, 1
%\bibitem{Moffatt1978}
%Moffatt H. K., 1978, Magnetic field generation in electrically conducting fluids. Cambridge, England, Cambridge University Press, 1978. 353 p.
%\bibitem{Park2017}
%Park K., 2017, Mon. Not. R. Astron. Soc., 472, 1628
%\bibitem{Park2020}
%Park K., 2020, ApJ, 898, 112
%\bibitem{Park2012a}
%Park K., Blackman E. G., 2012a, MNRAS, 419, 913
%\bibitem{Park2012b}
%Park K., Blackman E. G., 2012b, MNRAS, 423, 2120
%\bibitem{Park2014}
%Park K, 2014, MNRAS, 444, 3837–3844
%\bibitem{Pouquet}
%Pouquet A., Frisch U., Leorat J., 1976, Journal of Fluid Mechanics, 77, 321
%\bibitem{Tzeferacos}
%Tzeferacos, P., Rigby, A., Bott, A. et al. Laboratory evidence of dynamo amplification of magnetic fields in a turbulent plasma. Nat Commun 9, 591 (2018).
%\bibitem{Schrinner}
%Schrinner M., Rädler K.-H., Schmitt D., Rheinhardt M., Christensen U., 2005, Astronomische Nachrichten, 326, 245
%\bibitem{Semikoz}
%Semikoz V. B., 2004, arXiv e-prints, pp hep–ph/0403096
%\bibitem{Semikoz_Sokoloff}
%Semikoz V. B., Sokoloff D., 2005, A\&A, 433, L53
%\bibitem{Simon et al}
%Simon Cabanes, Nathanaël Schaeffer, Henri-Claude Nataf, 2014, Phys. Rev. Letter, 113, 184501
%\bibitem{Stefani}
%Stefani F., Giesecke A., Weber N., Weier T., 2016, Sol. Phys., 291, 2197
%\bibitem{Subramanian}
%Subramanian K., 2016, Reports on Progress in Physics, 79, 076901
%\bibitem{Tevzadze}
%Tevzadze A. G., Kisslinger L., Brand enburg A., Kahniashvili T., 2012, ApJ, 759, 54
%\bibitem{Warnecke}
%Warnecke J., Rheinhardt M., Tuomisto S., Käpylä P. J., Käpylä M. J., Brandenburg A., 2018, A\&A, 609, A51
%\bibitem{Yamazaki}
%Yamazaki D. G., Kajino T., Mathews G. J., Ichiki K., 2012, Phys. Rep., 517, 141
%\bibitem{Yoshizawa}
%Yoshizawa A., 2011, Hydrodynamic and Magnetohydrodynamic Turbulent Flows, 1998, Springer
%\bibitem{Yoshizawa_Itoh}
%Yoshizawa A., S. Itoh, K. Itoh, Plasma and Fluid Turbulence: Theory and Modelling, 2003, CRC press
%\bibitem{Yamazaki et al}
%Yamazaki, Dai G. and Kusakabe, Motohiko, 2012, Phys. Rev. D., 86, 123006



\end{thebibliography}
%\bibliographystyle{JHEP}
%\begin{thebibliography}{99}
%\bibitem{}[\protect\citeauthoryear{Author}{2012}]{Author2012}
%\bibitem{}[\protect\citeauthoryear{Others}{2013}]{Others2013}

%%%%%%%%%%%%%%%%%%%%%%%%%%%%%%%%%%%%%%%%%%%%%%%%%%

%%%%%%%%%%%%%%%%% APPENDICES %%%%%%%%%%%%%%%%%%%%%

\appendix
\section{Appendix}
For Eq.(\ref{forcing amplitude fk}), Biermann's effect is represented as
\begin{eqnarray}
{\bf f}_{mag}=\nabla \bigg(\frac{p_e}{n_e e}\bigg)=\frac{\nabla n_{e}\times \nabla p_{e}}{n_e^2e},
\label{forcing amplitude Biermann}
\end{eqnarray}
which is a typical example of  nonhelical magnetic forcing dynamo (NHMFD). Also, lepton-neutrino interaction produces the electromagnetic instability like below:
\begin{eqnarray}
{\bf f}_{mag}=-\frac{G_F}{\sqrt{2}|e|n_e}\sum_{\nu_a}c_{A}^a\bigg[(n_0^-+n_0^+)\,{\hat{\bf b}}\frac{\partial \delta n_{\nu_a}}{\partial t}+(N_0^-+N_0^+)\nabla(\hat{\bf {b}}\cdot {\delta {\bf j}_{\nu_a}})\bigg].
\label{forcing amplitude neutrino}
\end{eqnarray}
Its axial vector term is represented as ${\bf f}_{mag}=\alpha' {\bf B}$, where $\alpha'$ is \citep{Semikoz, Semikoz_Sokoloff}\footnote{Fermi constant $G_F=10^{-5}/m^2_p$($m_p$: proton mass); $c_{A}^a=\mp0.5$ (axial weak coupling, $a:$ electron, muon, tau; $(-)$: electron, $(+)$': muon or tau); $\delta n_{\nu_a}$: neutrino density asymmetry; $\delta {\bf j}_{\nu_a}$ (neutrino current asymmetry); $n_0^{\pm}\sim (|e|B/2\pi^2)T ln\,2$ is the lepton number density at Landau level. $\lambda^{\nu}_{fluid}\sim t$ is a scale of neutrino fluid inhomogeneity.}:
\begin{eqnarray}
\alpha'
%&=&\frac{G_F}{2\sqrt{2}|e| B}\sum_{\nu_a}c_{A}^{\nu_{a}}\bigg(\frac{n_0^-+n_0^+}{n_e}\bigg)\frac{\partial \delta n_{\nu_a}}{\partial t}\nonumber\\
\sim\frac{ln\, 2}{4\sqrt{2}\pi^2}\bigg(\frac{10^{-5}T}{m_p^2\lambda^{\nu}_{fluid}}\bigg)\frac{\delta n_{\nu}}{n_{\nu}}.
\label{neutrino_alpha}
\end{eqnarray}
{\bf These functions are not the same as our forcing function nor used in our simulation.  However, $\alpha'$ is the result of $\hat{\bf {b}}\cdot {\delta {\bf j}_{\nu_a}}$, which can produce helical magnetic field (HMFD).}\\

\noindent {\bf Biermann's battery effect and neutrino-lepton interaction are developed with the quantum electromagnetic process and neutrino interaction.} Their temporally and spatially inhomogeneous electromagnetic instabilities can induce the magnetic field and magnetic helicity. This generated magnetic field is amplified and transferred in the plasma system according to the MHD process. Their amplification processes are qualitatively consistent with the (magnetic) forcing dynamo with the modified EMF (${\bf \xi}={\bf U}\times {\bf B}+{\bf f}-\eta {\bf J}$, \cite{Park2012b}, and references  therein).\\
%The most importnat difference in `$\bf f$' in both cases is helicity ratio, which decides whether the growth of large-scale magnetic field or small-scale field.

% Fig.\ref{fig7a}, \ref{fig7b} for solar magnetic evolution were made with Eq.(\ref{Solar_poloidal_magnetic_field}), (\ref{Solar_toroidal_magnetic_field}) including the effect of planets near the sun ($\alpha$\&$\beta$). The results show that the solar magnetic period elevates from 16.31 to 21.74 years. And, Fig.\ref{fig8}, \ref{fig9} are for the validation of negative $\beta$ effect in Eq.(\ref{beta_derivation_helical2}). Kinetic helicity in small scale regime is larger than kinetic energy. Also, the sudden change of $\beta$ profile from a negative value to positive one near $t\sim 200$ implies the role of helical velocity field $\langle {\bf u}\cdot \nabla\times {\bf u}\rangle$ ($\rightarrow 0$) in the $\beta$ effect.

\begin{figure*}
\centering{
   \subfigure[]{
     \includegraphics[width=7.1 cm]{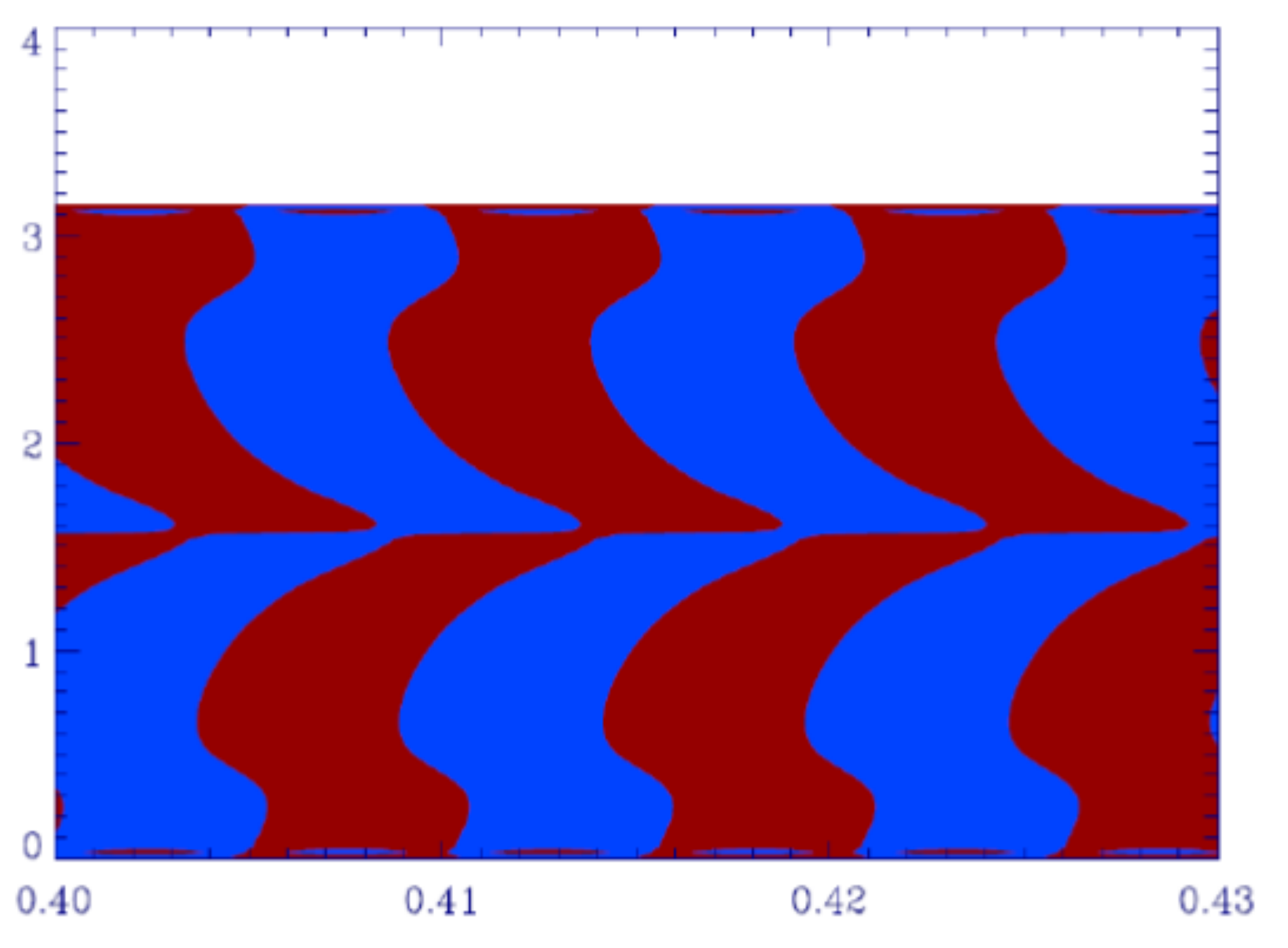}
      \label{fig7a}
}\hspace{-2 mm}
   \subfigure[]{
     \includegraphics[width=7.6 cm]{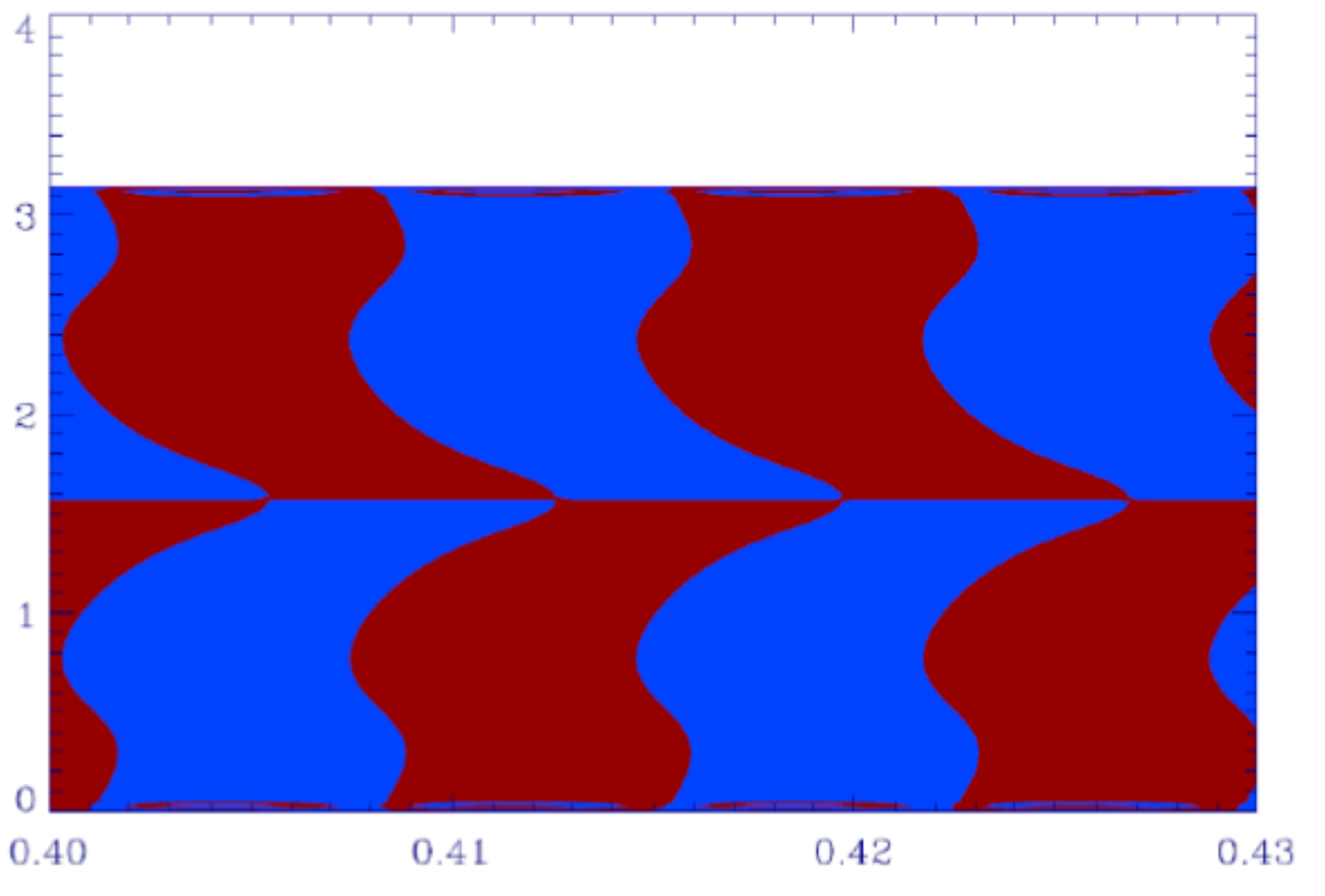}
     \label{fig7b}
   }
}
\caption{(a) 2D(azimuthal angle $\phi$ independent) simulation of Solar Magnetic field with Eq.~(\ref{Solar_poloidal_magnetic_field}), (\ref{Solar_toroidal_magnetic_field}). The simulation yields the period $\lambda=16.31$ years. (b) Tidal effect of solar planets is added to $\alpha$. As the tidal effect grows, the period  increases from $\lambda<22$ up to $\lambda=21.74$ years(Park 2021, not published). But in 1D, the period approaches in the opposite way.}
\end{figure*}

\begin{figure*}
\centering{
   \subfigure[]{
     \includegraphics[width=7.4 cm]{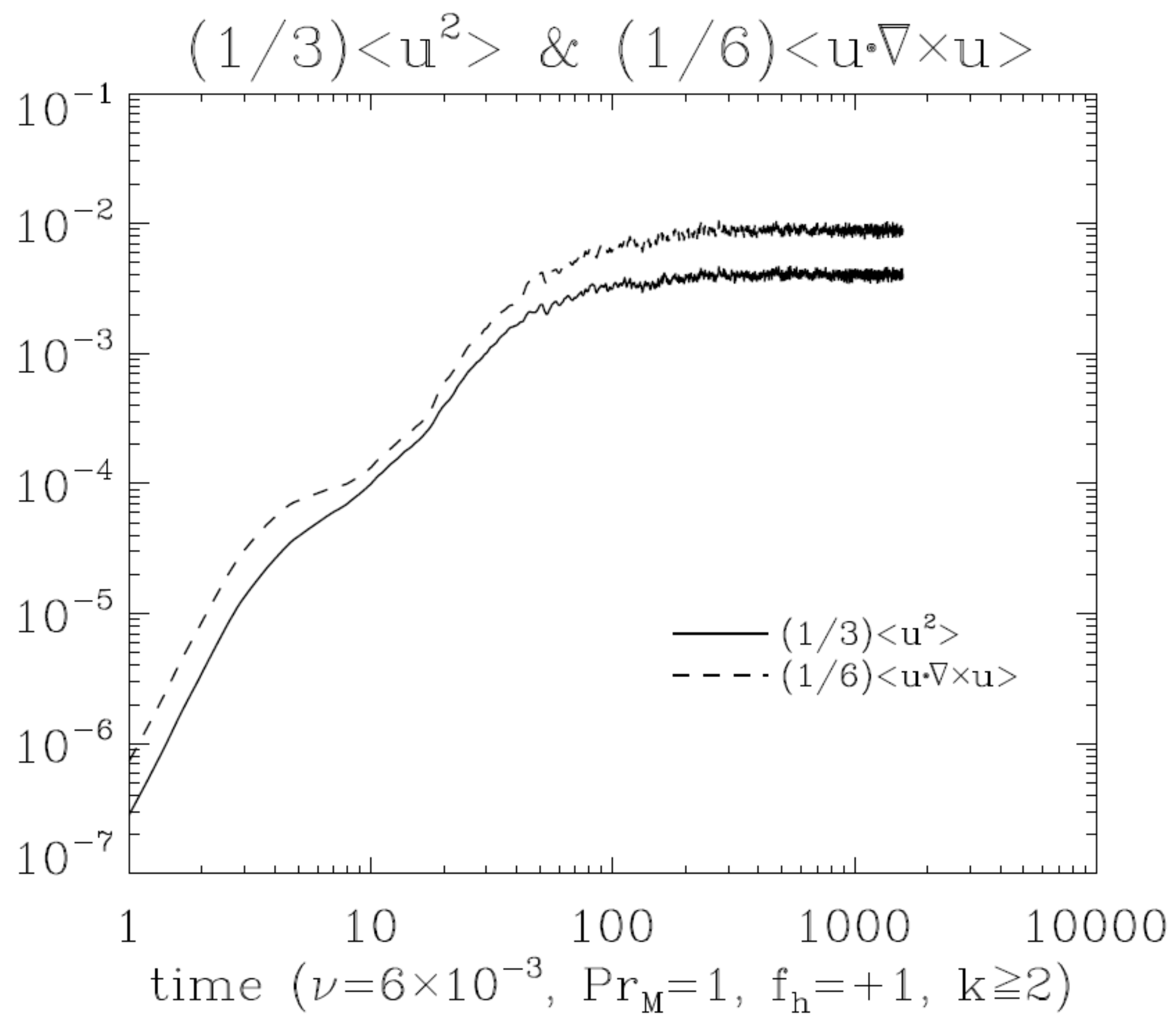}
     \label{fig8}
}\hspace{-5 mm}
   \subfigure[]{
     \includegraphics[width=7.68 cm]{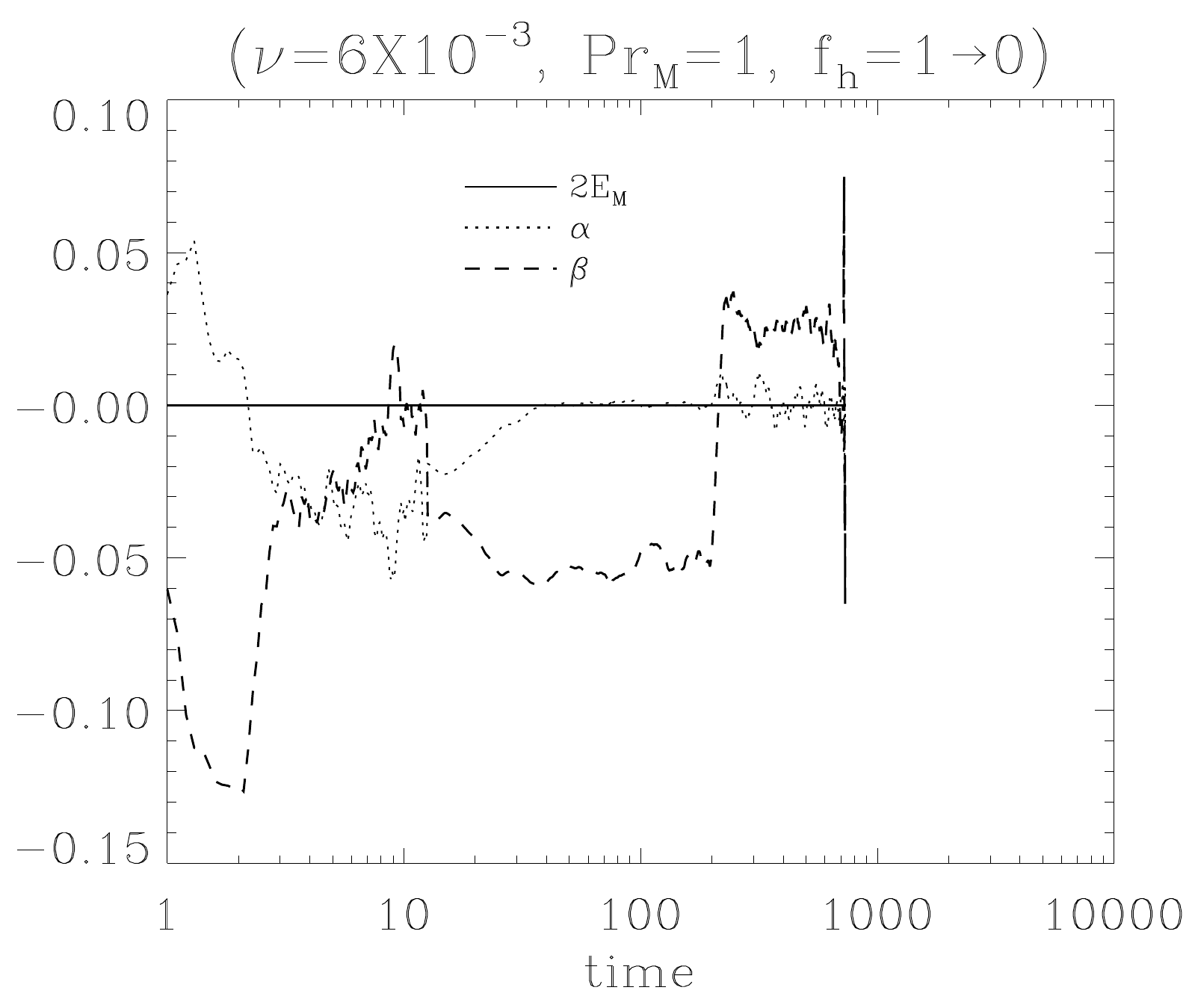}
     \label{fig9}
   }
}
\caption{(a) This plot is to compare the conventional $\beta$ effect from $E_V$ and that from $H_V$. All data for $k\geq 2$ are summed. (b) Helical kinetic forcing ($H_V\neq0$) is turned off at $t\sim 200$. And the system was continuously driven with nonhelical kinetic energy ($H_V=0)$ \cite{Park2020}.}
\end{figure*}

%\begin{figure*}
%\centering{
%{
%     \includegraphics[width=10.0 cm]{fig8}
%     \label{fig8Appen}
%   }
%}
%\caption{This plot is to compare the conventional $\beta$ effect from $E_V$ and that from $H_V$. All data for $k\geq 2$ are summed.}
%\end{figure*}
%
%
%
%\begin{figure*}
%\centering{
%{
%     \includegraphics[width=10.0 cm]{fig9}
%     \label{fig9}
%   }
%}
%\caption{Helical kinetic forcing ($H_V\neq0$) is turned off at $t\sim 200$ followed by nonhelical kinetic forcing ($H_V=0)$ \cite{Park2020}.}
%\end{figure*}
%%%%%%%%%%%%%%%%%%%%%%%%%%%%%%%%%%%%%%%%%%%%%%%%%%

% Don't change these lines
%%%%%%%%%%%%%%%%%%%%%\label{lastpage}

%%%%%%%\paragraph{Note added.} This is also a good position for notes added after the paper has been written.


\end{document}